%% file: ms.tex
\documentclass[fleqn,usenatbib]{mnras}

\usepackage{newtxtext,newtxmath}

\usepackage[T1]{fontenc}
\usepackage{hyperref}
\usepackage{breakurl}
\usepackage{ulem}
\DeclareRobustCommand{\VAN}[3]{#2}
\let\VANthebibliography\thebibliography
\def\thebibliography{\DeclareRobustCommand{\VAN}[3]{##3}\VANthebibliography}

\usepackage{graphicx}	

\title[Recent dust production events in TYC\,4209-1322-1]{Mid-infrared time-domain study of recent dust production events 
in the extreme debris disc of TYC\,4209-1322-1}

\author[A. Mo\'or et al.]{
Attila Mo\'or,$^{1,2}$\thanks{E-mail: moor.attila@csfk.org}
P\'eter \'Abrah\'am,$^{1,2,3}$
\'Agnes K\'osp\'al,$^{1,2,4,3}$
Kate Y.~L. Su,$^{5}$
George H. Rieke,$^{5,6}$
Kriszti\'an Vida,$^{1,2}$
\newauthor
Gianni Cataldi,$^{7,8}$
Attila B\'odi,$^{1,2,9}$
Zs\'ofia Bogn\'ar,$^{1,2,9}$
Borb\'ala Cseh,$^{1,2,10}$
G\'eza Cs\"ornyei,$^{1,2}$
N\'ora Egei,$^{1,2}$
\newauthor
Anik\'o Farkas,$^{1,2}$
Ott\'o Hanyecz,$^{1,2,11}$
Bernadett Ign\'acz,$^{1,2}$
Csilla Kalup,$^{1,2}$
R\'eka K\"onyves-T\'oth,$^{1,2}$
\newauthor
Levente Kriskovics,$^{1,2}$
L\'aszl\'o M\'esz\'aros,$^{1,2}$
Andr\'as P\'al,$^{1,2}$
Andr\'as Ordasi,$^{1,2}$
Kriszti\'an S\'arneczky,$^{1,2}$
B\'alint Seli,$^{1,2,12}$
\newauthor
\'Ad\'am S\'odor,$^{1,2,9}$
R\'obert Szak\'ats,$^{1,2}$
J\'ozsef Vink\'o$^{1,2,13}$
and Gabriella Zsidi$^{1,2}$
\\
$^{1}$Konkoly Observatory, Research Centre for Astronomy and Earth Sciences, E\"otv\"os Lor\'and Research Network (ELKH),\\ 
Konkoly-Thege Mikl\'os \'ut 15-17, H-1121 Budapest, Hungary \\
$^{2}$CSFK, MTA Centre of Excellence, Budapest, Konkoly Thege Mikl\'os \'ut 15-17., H-1121, Hungary\\
$^{3}$ELTE E\"otv\"os Lor\'and University, Institute of Physics, P\'azm\'any P\'eter s\'et\'any 1/A, H-1117 Budapest, Hungary\\
$^{4}$Max-Planck-Institut f\"ur Astronomie, K\"onigstuhl 17, D-69117 Heidelberg, Germany\\
$^{5}$Department of Astronomy/Steward Observatory, The University of  Arizona, Tucson, AZ 85721-0009, USA\\
$^{6}$Department of Planetary Sciences/Lunar \& Planetary Laboratory, The University of Arizona, 1629 E \\
 University Blvd, Tucson AZ 85721-0092, USA \\
$^{7}$National Astronomical Observatory of Japan, Osawa 2-21-1, Mitaka, Tokyo 181-8588, Japan \\
$^{8}$Department of Astronomy, Graduate School of Science, The University of Tokyo, Tokyo 113-0033, Japan\\
$^{9}$MTA CSFK Lend\"ulet Near-Field Cosmology Research Group\\
$^{10}$MTA-ELTE Lend{\"u}let "Momentum" Milky Way Research Group, Hungary\\
$^{11}$Faculty of Informatics, E\"otv\"os Lor\'and University, P\'azm\'any P\'eter s\'et\'any 1/A, H-1117 Budapest, Hungary\\
$^{12}$Department of Astronomy, E\"otv\"os Lor\'and University, P\'azm\'any P\'eter s\'et\'any 1/A, H-1117 Budapest, Hungary\\
$^{13}$Department of Optics and Quantum Electronics, University of Szeged, D\'om t\'er 9, H-6720 Szeged, Hungary \\
}

\date{Accepted XXX. Received YYY; in original form ZZZ}

\pubyear{2022}

\begin{document}
\label{firstpage}
\pagerange{\pageref{firstpage}--\pageref{lastpage}}
\maketitle

\begin{abstract}
Extreme debris discs are characterized by unusually strong mid-infrared 
excess emission, which often proves to be variable. 
The warm dust in these discs is of transient nature and is 
likely related to a recent giant collision occurring close to the star 
in the terrestrial region. Here we present the results of a 877\,d 
long, gap-free photometric monitoring performed by the {\sl Spitzer 
Space Telescope} of the recently discovered extreme debris disc around 
TYC\,4209-1322-1. By combining these observations with other time-domain optical 
and mid-infrared data, we explore the disc variability of the last four 
decades with particular emphasis on the last 12\,yr. During the latter 
interval the disc showed substantial changes, the most significant
was the brightening and subsequent fading between 2014 and 2018 as outlined
in {\sl WISE} data. The {\sl Spitzer} light curves outline the fading phase and a subsequent new brightening of the 
disc after 2018, revealing an additional flux modulation with a period of 
$\sim$39\,d on top of the long-term trend. We found that all these variations 
can be interpreted as the outcome of a giant collision that happened at an orbital 
radius of $\sim$0.3\,au sometime in 2014. Our analysis implies 
that a collision on a similar scale could have taken place around 2010, too. 
The fact that the disc was already peculiarly dust rich 40\,yr ago, as implied 
by {\sl IRAS} data, suggests that these dust production events belong to a chain 
of large impacts triggered by an earlier even more catastrophic collision.
\end{abstract}

\begin{keywords}
(stars:) circumstellar matter -- infrared: planetary systems -- stars: individual: TYC\,4209-1322-1
\end{keywords}



\section{Introduction}  \label{sec:intro}

We know that 20--30\% of main-sequence stars exhibit excess 
emission at infrared (IR) wavelengths, implying the presence of an optically thin 
circumstellar dust disc heated by the stellar light \citep{su2006,trilling2008,thureau2014,montesinos2016,sibthorpe2018}. 
The short lifetime of such dust grains implies that they are of second generation, 
produced via erosion of larger planetesimals, in most cases probably through their
steady state collisional grinding \citep{wyatt2008}. The majority of these debris discs, as we call 
the ensemble of solids of different sizes together, are comprised of cold material ($<$100\,K) 
residing at tens to hundreds of astronomical units (au) from the host star \citep{hughes2018}.   
These systems can be considered massive analogues of the Solar system's Kuiper-belt.
A few tens of per cent of such debris discs also display mid-IR excess \citep{kennedy2014}
indicating the presence of additional warmer dust material ($\gtrsim$150K), which 
is likely orbiting closer to the star. Furthermore, there are discs that contain only warm 
dust \citep[e.g.,][]{ballering2013}.
The observed warmer particles can be produced in situ in an inner planetesimal belt, but the sustenance
of this dust -- at least in part -- can also come from an outer, colder planetesimal belt
either via sublimation or disruption of comets transported from this outer reservoir 
or via grains migrated inward under the influence
of drag forces \citep[e.g.,][]{kennedy2015,marboeuf2016}. 

Mid-IR observations in recent decades have revealed a small but growing number of
debris discs -- mostly around FGK-type main-sequence stars -- that possess 
peculiarly large amounts of warm dust with temperatures 
higher than 300\,K \citep[e.g.,][]{song2005,melis2010,zuckerman2012,tajiri2020,melis2021,moor2021,vandenancker2021,higashio2022}. 
While in most known debris discs the fraction of the
stellar luminosity reradiated by the dust is lower than 10$^{-3}$, 
the dust material of these warm {\sl extreme debris discs} \citep[EDDs,][]{balog2009} 
intercepts at least one percent of the stellar light and thermally re-emits it
mostly in the mid-IR. These very high fractional luminosities suggest that these 
discs are at least partly optically thick for the stellar illumination.
In addition to their high dust content, EDDs show several features highly uncommon in 
normal debris systems. They tend to display significant photometric variability at 
3--5\,{\micron} on monthly to yearly timescales \citep{meng2014,meng2015,su2019,moor2021,su2022}, 
and have strong mid-IR solid state features, implying the presence of short-lived, 
small, submicron-sized crystalline dust particles \citep{olofsson2012}.
In contrast to typical debris discs, these characteristics cannot be explained by the slow, 
steady state collisional grinding
of an in situ planetesimal belt that started its evolution after the dispersal of the gas-rich initial 
primordial disc,
but instead point to a recent large transient dust production event -- likely related to a giant collision -- 
happening within 1--2\,au of the star \citep{jackson2012,su2019}.

While semi-annual photometric data 
from the Wide-field Infrared Survey Explorer \citep[{\sl WISE};][]{wright2010} satellite
allow us to get a picture of the annual variations of EDDs at 3.4 and 4.6\,{\micron}, 
dedicated monitoring projects with shorter cadences conducted with the 
{\sl Spitzer Space Telescope} \citep[{\sl Spitzer};][]{werner2004} made it possible to detect changes 
that happen over weeks or months. \citet{su2019} presented the 3.6 and 4.5\,{\micron} {\sl Spitzer} light curves 
of two EDDs, ID\,8 and P\,1121, both showing complex variations with shorter periodic (or quasi-periodic) 
modulations on top of longer-term trends. By attributing the dust production to recent collisional events, 
they found that the main features of the observed disc variability, including the periodic modulations, 
can be explained by the dynamical and collisional evolution of the ejected debris cloud. 
Their model allowed to constrain some important parameters of the event, such as the location of the collision and the 
size of the colliding bodies. While the light curve of P\,1121 is consistent with a giant impact 
that occurred prior 
to the start of the monitoring, the time-domain data of ID\,8 suggest at least two recent collisions 
at different times and locations \citep{su2019}.

Using the AllWISE infrared photometric catalogue \citep{cutri2013}, we have recently discovered a new 
EDD around TYC\,4209-1322-1 (hereafter TYC\,4209), a $\sim$275\,Myr old G1V type star 
\citep{moor2021} located at 270.8\,pc \citep{gaia2016,brown2020,cbj2020}. The star has an M3.5-type 
companion with a projected separation of 22\farcs2 \citep[$\sim$6000\,au,][]{moor2021}. 
TYC\,4209 exhibits excess in all four {\sl WISE} bands, between 3.4 and 22\,{\micron}, 
with a fractional luminosity 
of $L_{\rm disc} / L_{*} \sim$0.07, which is very high even among EDDs. Based on 
multi-epoch photometric data obtained with {\sl WISE} between 2010 and 2020, we found 
that this disc displayed significant variability at 3.4 and 4.6\,{\micron}, both on daily 
and yearly timescales \citep{moor2021}. By analysing the observed brightening events, we concluded that 
they cannot be explained by increasing the dust temperatures solely, but require the
formation of new dust material. 

In this work, we present a more detailed study of this disc and its mid-IR variability 
using time-domain photometry obtained with the Infrared Array Camera \citep[IRAC;][]{fazio2004} 
of {\sl Spitzer} at 3.6\,{\micron} (IRAC1) and 4.5\,{\micron} (IRAC2). 
Located close to the northern ecliptic pole at an ecliptic latitude of 87\fdg3, TYC\,4209 was in
the continuous viewing zone of {\sl Spitzer} 
allowing a gap-free 
monitoring over 877\,days. The optical variations of the object were also
monitored photometrically during the studied period and  
its spectrum between 2 and 3.8\,{\micron} was measured at three epochs using 
the IRTF/SpeX instrument.
We present the data reduction of these new observations in Sect.~\ref{sec:observations}.
The analysis of the stellar and disc variability and the determination of the fundamental disc parameters
are given in Sect.~\ref{sec:dataanalysis}. We interpret the observed disc variability in the framework 
of current models of giant impacts (Sect.~\ref{sec:discussion}). Finally the main outcomes of 
the study are summarized in Sect.~\ref{sec:summary}.
 
\section{Observations and data reduction} \label{sec:observations}

\subsection{Spitzer/IRAC multi-epoch imaging at 3.6 and 4.5\,{\micron}} \label{sec:spitzerdata}
IRAC observations of TYC\,4209 were performed in programs PID~13161 and PID~14071 
(PI: A. Mo\'or) in the post-cryogenic ('warm') phase of the {\sl Spitzer} mission. 
In the first program, 63 measurements were obtained between
2017 July 01 and 2019 March 14 using an average cadence of $\sim$10\,days. The second program 
started on 2019 April 06 and finished on 2019 November 25. In this, we conducted 48 observations 
utilizing a denser sampling with an average cadence of $\sim$5\,days. The longest gap 
in the time series ($\sim$23\,days) was between the end of the first program and the 
beginning of the second program. All observations were done in full-array observing mode\footnote{\url{https://irsa.ipac.caltech.edu/data/SPITZER/docs/irac/iracinstrumenthandbook/}} 
with exposure times of 1.2\,s using a 9-point random dither pattern with medium scale\footnote{\url{https://irsa.ipac.caltech.edu/data/SPITZER/docs/irac/calibrationfiles/dither/}}. 

\begin{figure} 
\begin{center}
\includegraphics[scale=.44,angle=0]{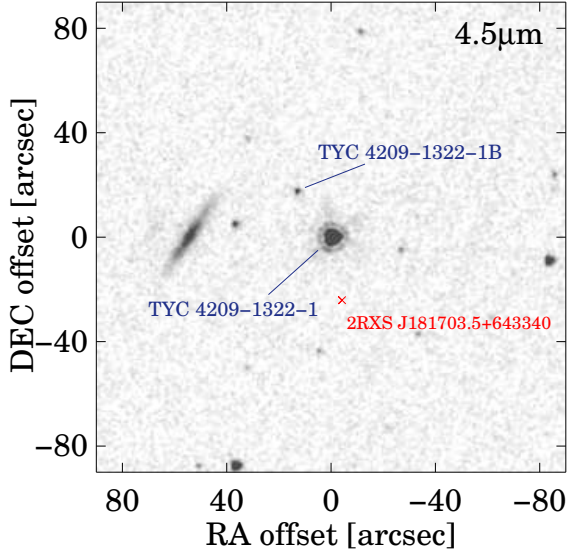}
\caption{ {Field of TYC\,4209 at 4.5\,{\micron} with {\sl Spitzer} (based on data obtained 
in AOR\,63191296).   
}
\label{fig:chart}
}
\end{center}
\end{figure}

In our data reduction, we used the datasets produced by the
Spitzer Science Center (SSC) with the IRAC pipeline
version of S19.2.0. These contain both processed images 
for the nine dither positions (Corrected
Basic Calibrated Data, CBCDs) and mosaic images compiled 
from them. Figure~\ref{fig:chart} shows the immediate vicinity 
of TYC\,4209 at 4.5\,{\micron}. As this mosaic image demonstrates,
it appears as a well-separated, bright point source.    
The photometric analysis was performed on the individual CBCD frames.
We used aperture photometry with a radius of 
3\,pixels (the native pixel size is 1\farcs22) 
to extract flux densities from these data. 
To estimate the centroid position of our target
we employed a first moment box centroider algorithm proposed by the SSC 
(\texttt{box\_centroider}\footnote{\url{https://irsa.ipac.caltech.edu/data/SPITZER/docs/dataanalysistools/tools/contributed/irac/box_centroider/}}). 
The background was measured in a sky annulus between radii of 12 and 20 pixels using an iterative 
sigma-clipping method with a clipping threshold of 3$\sigma$. 
For the aperture photometry we utilized the \texttt{daophot} package of the 
Interactive Data Language (IDL) astronomy library \citep{landsman1995}.
The obtained photometry was corrected both for the pixel phase effect and for array location dependent response
utilizing the \texttt{irac\_aphot\_corr}\footnote{\url{https://irsa.ipac.caltech.edu/data/SPITZER/docs/dataanalysistools/tools/contributed/irac/iracaphotcorr/}} 
task provided by the SSC.
We then applied aperture correction using the appropriate correction factors taken 
from the IRAC Instrument Handbook v3.0.1\footnote{\url{https://irsa.ipac.caltech.edu/data/SPITZER/docs/irac/iracinstrumenthandbook/}}.
The final flux density and its uncertainty were derived by calculating the centre and dispersion of the 
distribution of the nine individual photometric data points obtained at the different dither positions 
using bisquare weighting 
(implemented in the \texttt{biweight\_mean} IDL task\footnote{\url{https://idlastro.gsfc.nasa.gov/ftp/pro/robust/biweight_mean.pro}}).
The photometric results are listed in Table~\ref{tab:iractable}.


\begin{figure} 
\begin{center}
\includegraphics[scale=.44,angle=0]{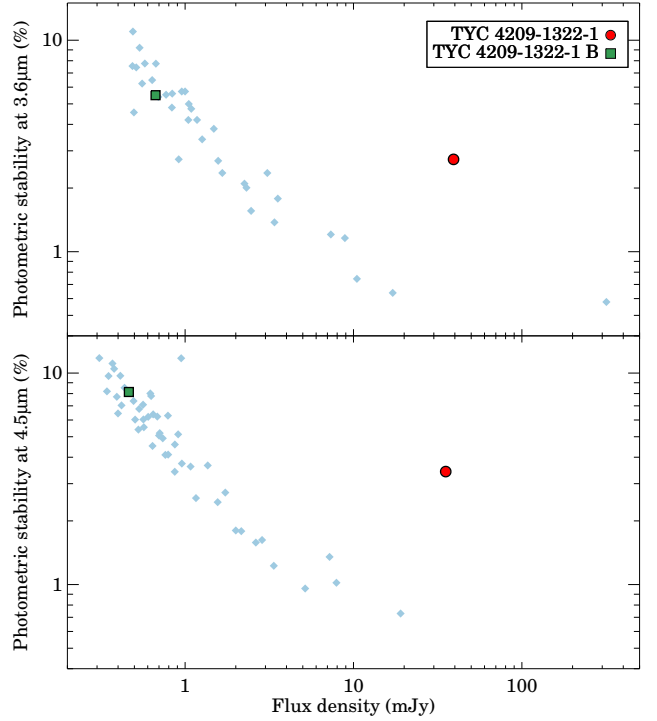}
\caption{ {Photometric stability of TYC\,4209 and other sources in its 
vicinity at 3.6 and 4.5\,{\micron} as a function of their flux density in the specific
band during the {\sl Spitzer} measurement programme.   
}
\label{fig:photstability}
}
\end{center}
\end{figure}


To evaluate the photometric stability during the monitoring program, we constructed 
the 3.6 and 4.5\,{\micron} light curves of nearby objects covered by our observations. 
Due to the optical design of the IRAC instrument, the IRAC1 and IRAC2 arrays were operated 
simultaneously providing two nearly adjacent 5\farcm2$\times$5\farcm2 field of views 
with one covering the target. 
The nearest edges of the images are separated by $\sim$1\farcm52. As the field of view rotated from 
epoch to epoch, only sources in the immediate vicinity of TYC\,4209 were measured 
at all times. We considered only those objects for which at least 10 measurements 
were made with a difference of at least two years between the first and last point. 
The lists of sources are only partially identical in the two bands.
Photometry of these additional targets was obtained using the same procedure as described 
above for TYC\,4209, the photometric stability was calculated as the standard deviation of 
the normalized flux densities.
Figure~\ref{fig:photstability} shows the derived photometric stabilities as a function 
of the flux level for both IRAC channels. At the flux level of TYC\,4209, these results 
suggest a stability better than 0.7\% and 1\% at 3.6 and 4.5\,{\micron}, respectively. 
TYC\,4209 clearly stands out from the trend in both bands, proving that it showed significant 
changes during the monitoring period. In addition to our main target, at 4.5\,{\micron} there 
is another outlying object, WISE J181805.36+642954.6, that is classified as a W\,UMa type 
eclipsing binary based on its ASAS-SN optical light curve \citep{jayasinghe2018}. 
It is therefore not surprising that this system also proves to be variable 
in the mid-IR.

\input{tab1.tex}

Similarly to TYC\,4209, its companion was also measured at each epoch.
As Fig.~\ref{fig:photstability} demonstrates, data of TYC\,4209\,B do not imply any 
significant changes over the observed period. By computing the weighted average of all 
available 111 photometric data points, for this star we obtained flux densities of 
0.66$\pm$0.02\,mJy and 0.47$\pm$0.01\,mJy at 3.6 and 4.5\,{\micron}, 
respectively. 
The quoted uncertainties are calculated by adding quadratically the measurement 
and absolute calibration uncertainties, 
where the latter component is assumed to be 
2\%\footnote{\url{https://irsa.ipac.caltech.edu/data/SPITZER/docs/irac/warmimgcharacteristics/}}.
In \citet{moor2021}, we derived an effective temperature of 3230$\pm55$\,K for TYC\,4209\,B. 
Assuming that, like the primary component, it has solar metallicity and adopting a 
$\log{g}$ value of 5.0\,dex, we fitted a NextGen atmosphere model to the Gaia\,EDR3 and 2MASS 
photometric data of the star. This model predicts photospheric flux densities of 0.68$\pm$0.01\,mJy
and 0.45$\pm$0.01\,mJy in the IRAC\,1 and IRAC\,2 bands, respectively, which are consistent with 
the measured fluxes, i.e., TYC\,4209\,B shows no 
excess in these bands.


\begin{figure}
\begin{center}
\includegraphics[scale=.45,angle=0]{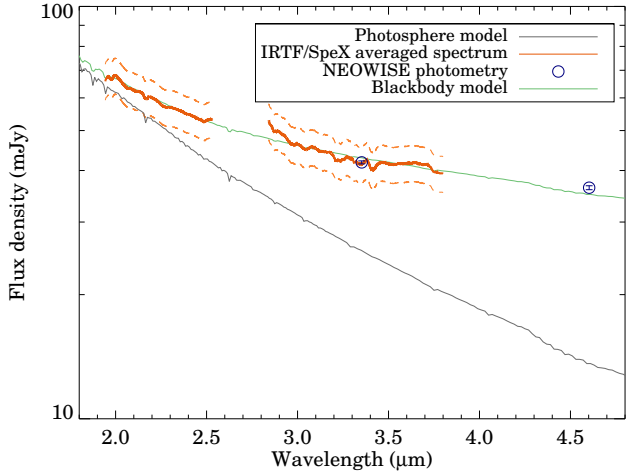}
\caption{ {The SpeX/IRTF spectrum of TYC\,4209. The plotted spectrum is obtained 
as the average of three individual spectra (see Sect.~\ref{sec:irtfspectra}). 
Flux uncertainties are marked by dashed orange lines, mainly 
dominated by the 10\% systematic instrumental errors. 
The stellar photosphere model and a blackbody model (with T = 1040\,K) fitted to the 
measured spectrum (Sect.~\ref{sec:diskproperties}) are also plotted.  
}
\label{fig:irtfspectra}
}
\end{center}
\end{figure}

\subsection{WISE multi-epoch photometry}  \label{sec:wisedata}
The discovery of the disc around TYC\,4209 and the first study of its infrared variations 
were based on data obtained with the {\sl WISE} satellite \citep{moor2021}. Since the start of its mission in 2010, 
{\sl WISE} has collected data on TYC 4209 in 18 observing windows, with a time shift of $\sim$6\,months between 
them, except for a hibernation period of the telescope, between 2011 and 2013, when no observations were performed.  
While in the first window, during the cryogenic phase, measurements were made in all four photometric bands 
available ($W1$, $W2$, $W3$, and $W4$ at 3.4, 4.6, 12, and 22\,{\micron}, respectively), 
in the subsequent windows, observations were restricted to the two shortest wavelength bands. 
Thanks to its advantageous position close to the ecliptic pole, for TYC\,4209 the observation windows were quite 
long (with widths of 17--40\,days), allowing the telescope to obtain 40--300 individual single 
exposures in each window. In \citet{moor2021}, after discarding bad quality exposures, 
 we computed the average of the remaining data points in each observing window between 2010 and 2019. 
The continuation of the NEOWISE Reactivation program \citep{mainzer2014} in 2020--21 allowed us to derive 
3.4 and 4.6\,{\micron} photometry in four additional epochs\footnote{\url{https://www.ipac.caltech.edu/doi/irsa/10.26131/IRSA144}}. These were calculated using the same 
method as employed in our previous work.   


\begin{figure*} 
\begin{center}
\includegraphics[scale=.45,angle=0]{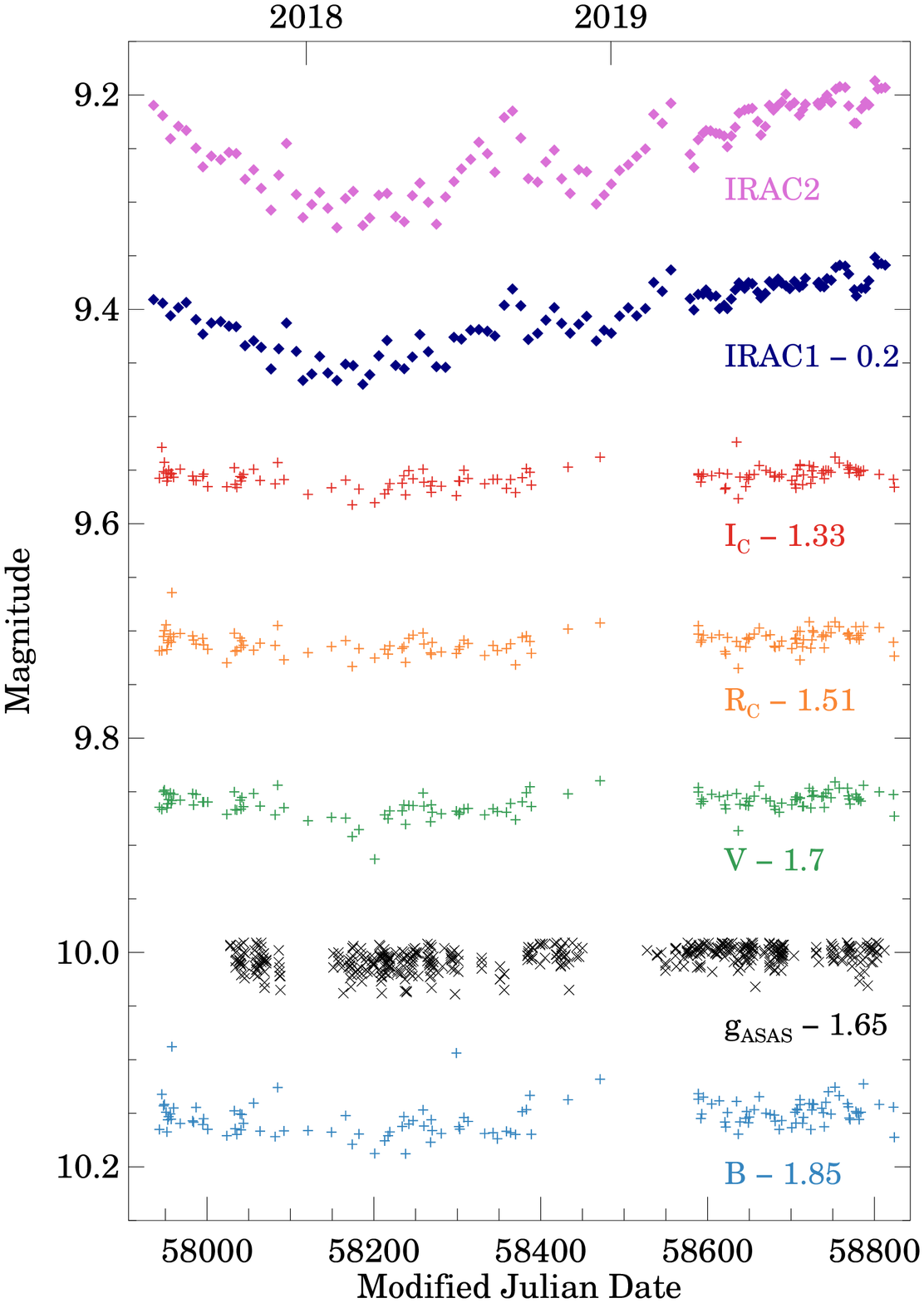}
\includegraphics[scale=.45,angle=0]{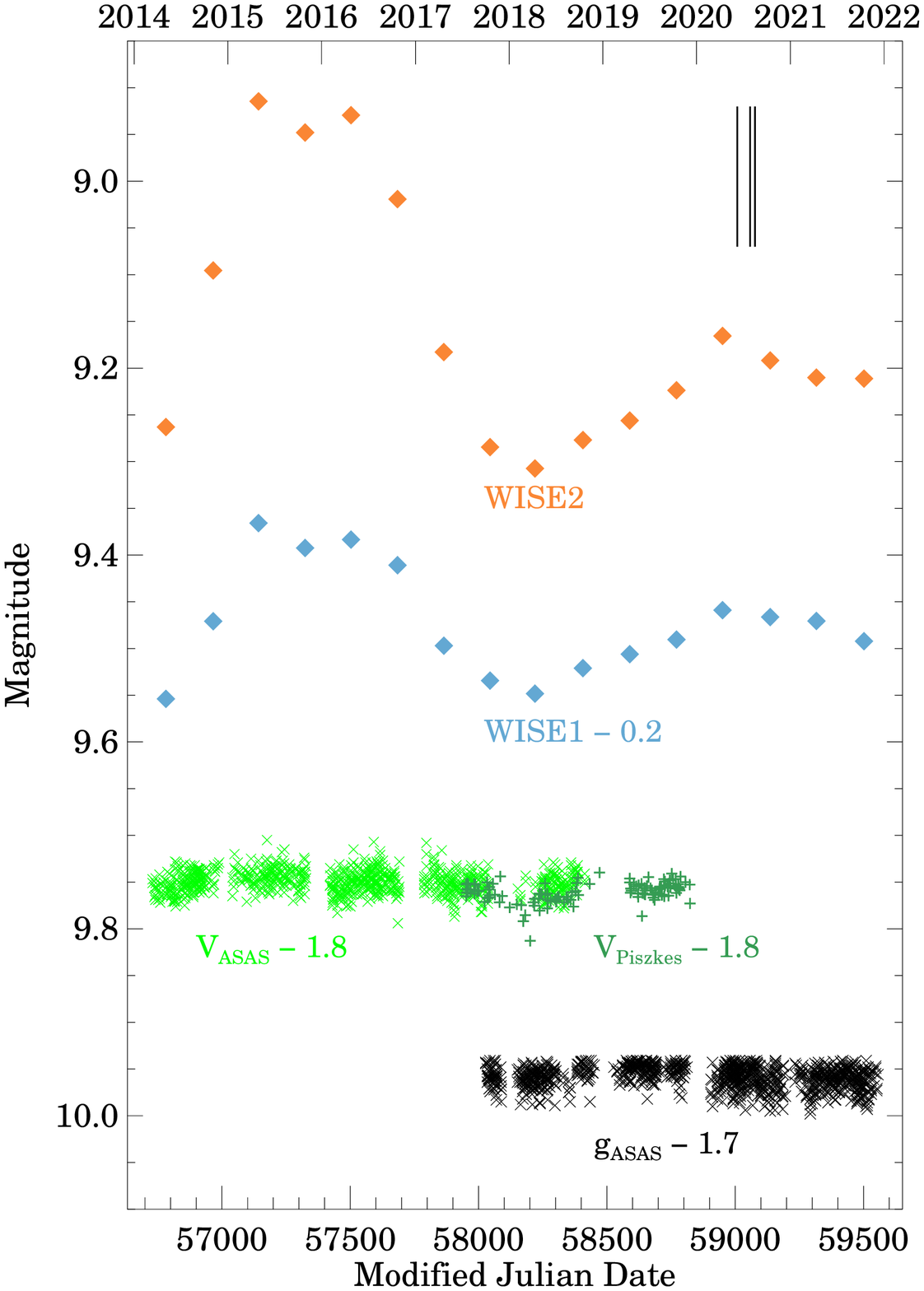}
\caption{ {Optical and infrared light curves of TYC\,4209, during the {\sl Spitzer} monitoring campaign, 
between 2017 and 2019 (left panel) and during the NEOWISE Reactivation mission between 2014 and 
2022 (right panel). For clarity, the light curves are shifted along the y axis by the 
indicated magnitudes. The utilized photometric data sets are described in Sect.~\ref{sec:observations}. 
Black vertical lines in the right panel mark the epochs of our IRTF/SpeX observations.  
}
\label{fig:lightcurves}
}
\end{center}
\end{figure*}

\subsection{IRTF/SpeX mid-IR spectroscopy} \label{sec:irtfspectra} 
We observed TYC\,4209 three times on 2020 June 18, July 28, and August 16 with the NASA IRTF 
telescope using the SpeX instrument \citep{rayner2003}. We selected nearby A0V star, HIP\,94140 
(B=6.26 and V=6.26) and HIP\,91315 (B=5.69 and V=5.74) as telluric standards for atmospheric 
correction. Observations were made with the LXD\_long mode with a slit of 0\farcs8$\times$15\arcsec, 
providing spectra covering 1.98--5.3\,{\micron} with $R$=2500. The sky conditions for all three nights 
were stable with a typical seeing of $\sim$0\farcs9. All observations were made using the ABBA 
nod patterns to remove telescope and sky background. An integration time of 60\,s was used for 8 
cycles on the target, resulting in a total on-target integration time of 960\,s for each of the epochs. 
Internal arc lamps were used for wavelength calibration, along with sky lines at the longer wavelengths. 
Spectral flat fields were taken with internal lamps, as described in \citet{rayner2003}. 

SpeX data were reduced using the SpeXtool software package \citep{cushing2004} following the standard 
reduction procedure \citep{vacca2003}. The target is too faint to have good signal-to-noise spectra 
at the longest wavelengths (order 4); we then trimmed any data points longwards of 3.8 $\mu$m. For each 
epoch, multi-cycle spectra were then combined using the robust weighted mean method. We found no 
convincing flux variability (above $\sim$10\% level) among the three epochs. We then combined the 
three epoch spectra by weighted average and smoothed to $R$=200 for assessing the general shape 
of the infrared excess as shown in Figure~\ref{fig:irtfspectra}. 
All three IRTF observations were conducted between the two observing windows of the NEOWISE 
Reactivation project in 2020. The $W1$ and $W2$ band fluxes of the source show only small changes 
($<$3\%) between these two windows. Figure~\ref{fig:irtfspectra} displays averages of the 
two $W1$ and $W2$ 
observations. As the plot demonstrates, the average of the $W1$ measurements coincides very 
well with the averaged IRTF spectrum indicating that the absolute calibration of the latter 
is of good quality.

\subsection{Optical monitoring} \label{sec:opticalmonitoring}
The emission measured in the IRAC bands comes partly from the dust disc and partly from 
the photosphere of the star. It is therefore also important to monitor the variations 
in the stellar brightness over the studied period. For this purpose, we conducted
optical photometric observations using Bessel $BV(RI)_{\rm C}$ filters
at the Piszk\'estet\H o Mountain Station 
of Konkoly Observatory (Hungary) using the 60/90/180\,cm Schmidt telescope equipped with a
4096$\times$4096 pixel Apogee Alta U16 CCD camera.  
The camera had a pixel scale of 1$\farcs$03 and a field of view of 1$\fdg$17$\times$1$\fdg$17.
Over the period covered by our {\sl Spitzer} monitoring project we obtained 127 imaging blocks 
during 123\,nights. The average cadence was 6.8\,days. In each imaging block, 
typically five images were taken with each filter.
We reduced the images in IDL following the standard processing steps of bias and dark 
subtraction and flat-fielding. We calculated aperture photometry for the target and several 
comparison stars using an aperture radius of 5 pixels and sky annulus between 20 and 40 pixels. 
We used on average 170 comparison stars within a half degree radius of TYC\,4209. We took 
standard magnitudes for the comparison stars from the APASS9 catalogue \citep{henden2015} and converted 
the Sloan $g'r'i'$ magnitudes to $R_{\rm C}I_{\rm C}$ by spline interpolating in the comparison 
stars' SEDs for the effective wavelength of these filters. We obtained the calibration factor by 
fitting a linear colour term for the $V-R_{\rm C}$ colour and iterating until the photometry of the 
science target converged. 
Typical rms scatter of the photometry are 0.009 mag in the $B$ band and 
0.006 mag in the $VR_{\rm C}I_{\rm C}$ bands.
In the data analysis we utilized the daily average of the obtained magnitudes in each filter. 
The measured $B, V, R_C, I_C$ light curves together with the IRAC1 and IRAC2 photometric data (in magnitudes) 
are displayed in Figure~\ref{fig:lightcurves} (left).

\medskip

TYC\,4209 was covered by the All-Sky Automated Survey for 
Supernovae \citep[ASAS-SN,][]{shappee2014,kochanek2017} photometry survey, 
providing $V$ band data between 2012 April and 2018 October, and $g$ band 
observations since 2017 October. The latter data (see Fig.~\ref{fig:lightcurves}, left) 
similarly to our observations from Piszk\'estet\H o, 
cover well the period of the {\sl Spitzer} IR monitoring with a median cadence of 1.75\,days.
The V-band measurements provide information about stellar variability 
in the first five years of the NEOWISE Reactivation mission, 
allowing its comparison 
with the changes seen in the $W1$ and $W2$ bands in that period (Fig.~\ref{fig:lightcurves}, right).

\section{Data analysis} \label{sec:dataanalysis}

\subsection{Optical variability} \label{sec:opticalvariability}
Using the available optical light curves (Sect.~\ref{sec:opticalmonitoring}), we explored the 
stellar variability and investigated its possible contribution to the observed changes in 
the mid-infrared bands. As Figure~\ref{fig:lightcurves} (left) demonstrates, during the {\sl Spitzer} 
monitoring the system exhibited much larger amplitude changes at 3.6 and 4.5{\micron} than in the 
optical bands. The standard deviation of the $B$, $V$, $R_C$, and $I_C$ light curves are 
0.013, 0.010, 0.010, and 0.009\,mag, respectively, while for the corresponding part of 
the ASAS $g$-band photometry, 0\fm009 is obtained. This indicates that the stellar brightness 
is stable within $\sim$1\%. 
The contribution of the star to the total flux is 56--63\% and 38--43\% at 3.6 and 4.5{\micron}, respectively.
So even if the variations of the star at 3.6 and 4.5{\micron} are similar in amplitudes to those in 
the optical bands, they are negligible compared to the mid-infrared changes observed with amplitudes 
of $\sim$11\% and $\sim$13\%, respectively. 
As Figure~\ref{fig:lightcurves} (right) 
shows, the same is particularly true for the variations observed between 2014 and 2018: the 
changes in the star's $V$ band photometry ($\sigma_{V_\mathrm{ASAS}}$=0\fm013) are dwarfed by 
the ones seen in the $W1$ and $W2$ bands. 

In addition to ground-based optical data, space photometry is also available for TYC 4209 thanks 
to the Gaia and the Transiting Exoplanet Survey Satellite \citep[TESS,][]{ricker2015} space telescopes.
TYC\,4209 was observed in 10 sectors (each $\sim$27\,days long) during Year~2 of the TESS mission 
in the time range between 2019 August 15 (MJD 58711.4) and 2020 July 4 (MJD 59035.1), the first 65 days 
of this data series overlap with the {\sl Spitzer} light curves. By analysing the whole TESS 
data set, we revealed a modulation with a period of 
5.07\,days and an amplitude of 0.0032\,magnitude, attributed to starspots and stellar 
rotation \citep{moor2021}. We found some weaker peaks around the dominant one in 
the Fourier spectrum and interpreted them as signs of differential rotation. 
The effect of rotational modulation can also be detected in the ASAS light curves.
By calculating the generalized Lomb--Scargle (GLS) periodograms of the $V$- and $g$-band 
data \citep{zechmeister2009} using 
the IDL implementation of the publicly available code\footnote{\url{https://github.com/mzechmeister/GLS}}, 
we found significant peaks at $\sim$5.13\,d and $\sim$5.04\,d, with false alarm probabilities (FAPs)
of 1$\times$10$^{-4}$ and 8$\times$10$^{-9}$, and with amplitudes of 
0.0035 and 0.0042\,mag,
respectively. There is an additional 29.5-day period in the $V$ light curve, but we also found that period 
in the ASAS data of most surrounding stars, implying that it is probably 
an instrumental artifact caused by the lunar phase, as it is identical to the length of the 
synodic month. 
According to Gaia~EDR3 \citep{brown2020}, TYC\,4209 
was observed photometrically at 355 epochs in the $G$ band and at 41 epochs in the 
$G_\mathrm{BP}$ and $G_\mathrm{RP}$ bands. These observations were conducted between 
2014 July 25 and 2017 May 28, during a period when the object significantly 
brightened and then faded  in the mid-IR (Fig.~\ref{fig:lightcurves}, right).  Although individual data points 
are not available, the relative standard deviation of the multi-epoch Gaia photometric data can 
be calculated as 100$\times \sqrt{N_\mathrm{obs}} \frac{\sigma_f}{f}$ based on the number of measurements 
($N_\mathrm{obs}$) and the flux ($f$) and flux error ($\sigma_f$) listed in the EDR3 catalogue. 
This yields relative standard deviations of 0.77, 0.83, and 0.48\% for the $G$, $G_\mathrm{BP}$ and 
$G_\mathrm{RP}$ data sets, respectively. This result also indicates that the optical brightness of the 
star was very stable over the given period.

\begin{figure} 
\begin{center}
\includegraphics[scale=.50,angle=0]{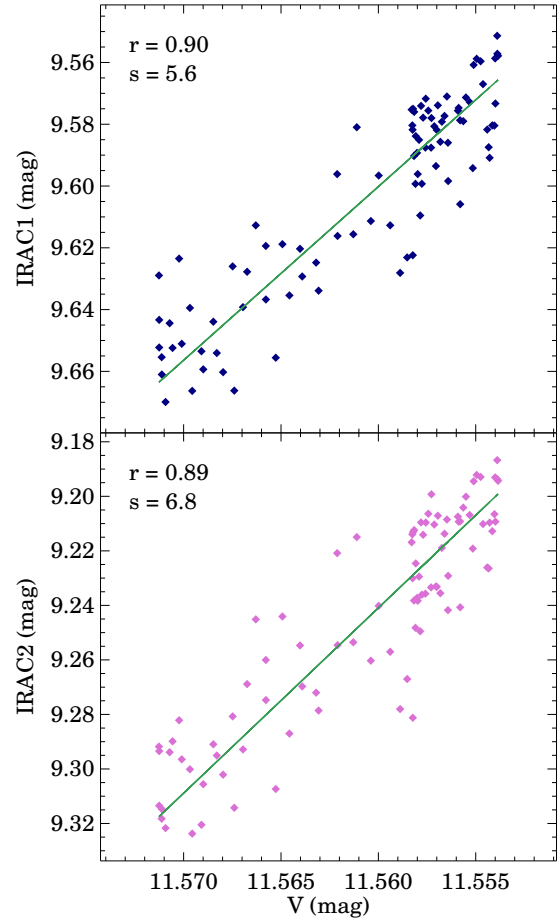}
\caption{ {Scatter plots of the measured V and IRAC1 (top), IRAC2 band (bottom) photometric 
data with the best fit lines. Values of the Pearson correlation coefficients ($r$) and 
the derived slopes ($s$) of the fitted lines are also shown.    
}
\label{fig:scatterplots}
}
\end{center}
\end{figure}

Interestingly, the optical light curves between 2017 and 2019 (Fig.~\ref{fig:lightcurves}, left) 
show a trend that follows the changes observed in the mid-infrared wavelengths, albeit with 
very low amplitude. We applied the locally weighted scatterplot smoothing (LOWESS) algorithm to 
the $V$-band light curve to find the trend and then interpolated it to 
the epochs of the {\sl Spitzer} measurements.
In Fig.~\ref{fig:scatterplots}, we display scatter plots of  
IRAC1 and IRAC2 photometry against this $V$-band trend.   
In the period between MJD 58400 and 58585, practically there is no $V$-band photometry available, making the 
trend estimate unreliable,  
therefore {\sl Spitzer} measurements from this time range have not been used in the plot. 
The figure indicates a correlation between the mid-infrared and the $V$-band data. 
The Pearson correlation coefficient between the IRAC1 and the $V$-band photometry is 
$r_\mathrm{IRAC1-V} = 0.9$, while for the IRAC2 and $V$ data pairs we obtained 
$r_\mathrm{IRAC2-V} = 0.89$. Comparing the other optical band measurements with the 
{\sl Spitzer} data also indicates significant correlations. Considering IRAC1 observations, 
the correlation coefficients between them and data 
measured with the $B$, $R_\mathrm{C}$, and $I_\mathrm{C}$ filters are 
0.91, 0.89, and 0.86, respectively, while for IRAC2, 0.88, 0.89, and 0.88.  
By applying linear regression for the 
data pairs using an ordinary least squares bisector method \citep{isobe1990},
we obtained slope values of 4.7$\pm$0.4, 5.6$\pm$0.4, 7.5$\pm$0.6, and 8.3$\pm$0.8 
for IRAC1 versus $B$, $V$, $R_\mathrm{C}$, and $I_\mathrm{C}$ 
data, and 5.6$\pm$0.5, 6.8$\pm$0.6, 9.1$\pm$0.8, and 10.1$\pm$0.9
for the relations between IRAC2 and the same optical photometry. 
 The mid-IR brightness variations of the system are likely related to the changes in 
 the emitting area  
  of the dust disc \citep[][Sect.~\ref{sec:colortemperatures}]{moor2021}. Based on the above relations, the 
 optical brightness variations of the system are correlated with this, and 
 when the disc is brighter in the mid-IR, the system becomes brighter and bluer at optical wavelengths.  
 A possible explanation is that these optical changes are related to the scattering of starlight 
 by the fresh dust particles. 
 However, the observed 
 relationship does not seem to be as strong in all periods. Comparing the ASAS $V$ 
 photometry with the {\sl WISE} half-annual data yields a somewhat weaker correlation with 
 coefficients of $r_\mathrm{W1-V} = 0.69$, and $r_\mathrm{W2-V} = 0.68$. 
 In addition, the resulting slopes are much larger, 15.6$\pm$2.6 for $W1$ and $V$ data pairs and 
 29.4$\pm$4.8 for $W2$ and $V$ data. Although these WISE bandpasses are not identical 
 to the IRAC bandpasses, the $W1$ ($W2$) filter is very similar to the 
 IRAC1 filter (IRAC2). So the difference in the slopes may not be due to the difference in 
 the filters, but to the fact that the relationship 
 in the earlier 2014--2018 period was different from the period covered by 
 the {\sl Spitzer} measurements.    
If the observed effect is indeed related to starlight scattered by 
dust grains, then it suggests the presence of different dust 
populations in the two periods (Sect.~\ref{sec:yearlyvar}).

\begin{figure*} 
\begin{center}
\includegraphics[scale=.50,angle=0]{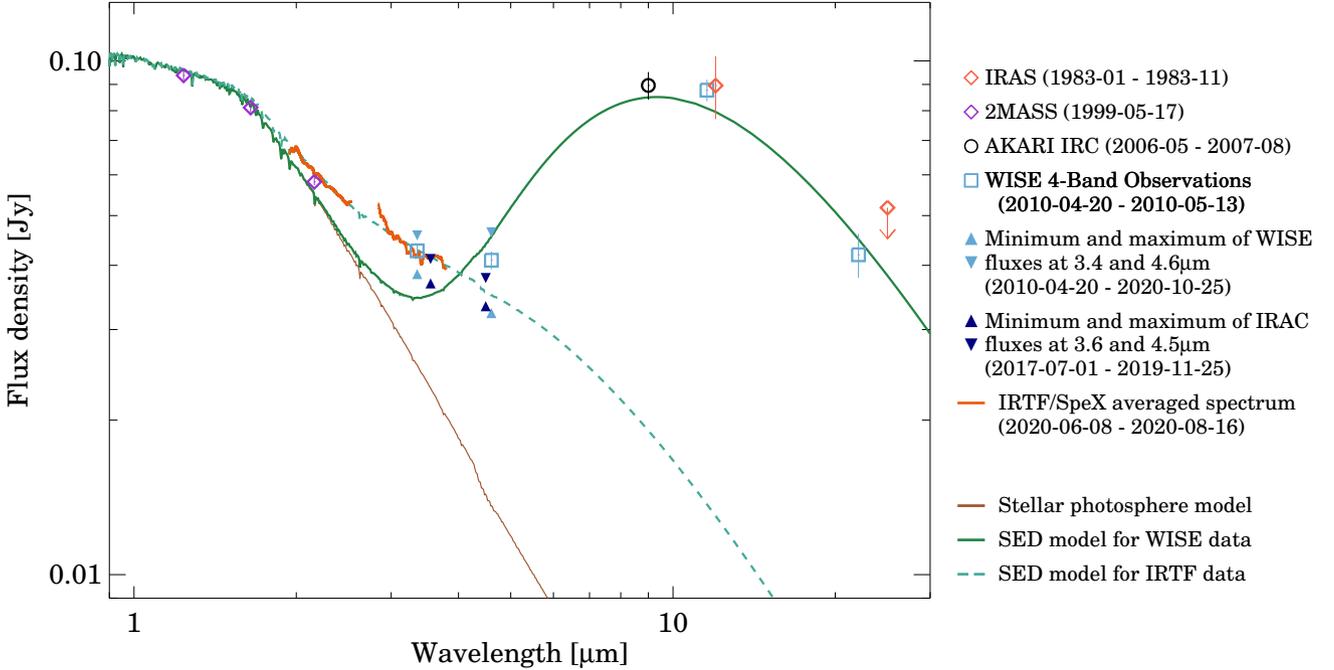}
\caption{ {Spectral energy distribution of TYC\,4209. 
The green solid and dashed curves show the best fitting SED models to the {\sl WISE} four band data
measured in 2010 \citep{moor2021} and to the IRTF spectrum in 2020 (Sect.~\ref{sec:irtfspectra}), respectively. 
}
\label{fig:sed}
}
\end{center}
\end{figure*}

\subsection{General disc properties} \label{sec:diskproperties}
Figure~\ref{fig:sed} shows the spectral energy distribution (SED) 
of TYC\,4209 with all available near- and mid-infrared photometric and spectroscopic 
observations. Comparing the best fitting Kurucz model 
of the stellar photosphere \citep[taken from][]{moor2021} with these data, indicates that 
the object exhibits excess at all wavelengths longward of  
2\,{\micron}. To estimate the basic disc properties, in \citet{moor2021} we 
fitted a single temperature blackbody model (green solid curve in Fig.~\ref{fig:sed}) 
to the excess measured 
in the four {\sl WISE} bands during the cryogenic phase of the 
mission (blue open squares in Fig.~\ref{fig:sed}). This fit yielded a characteristic dust temperature of 
$\sim$530\,K and a fractional luminosity of $\sim$0.07.
The applied model is clearly oversimplified. Firstly, most EDDs 
show strong solid state emission features. TYC\,4209 does not have any mid-IR spectral
information, therefore,  we cannot assess how such features might affect the shape of the SED in 
our case. Secondly, the single temperature blackbody model significantly underestimates the flux data at 
3.4\,{\micron}, suggesting the presence of 
additional hotter dust particles in the disc. Measurements obtained at other epochs 
also support the latter hypothesis. {\sl WISE} and IRAC data point pairs 
indicate dust colour temperatures between 760 and 1000\,K (Sect.~\ref{sec:colortemperatures}), while the excess measured by the 
IRTF spectrum can be fitted very well with a 1040$\pm$10\,K blackbody model that 
has a fractional luminosity of 0.04 (Figure~\ref{fig:irtfspectra} and \ref{fig:sed}). 

The minima and maxima of the {\sl WISE} $W1$ and $W2$ band flux data in Fig.~\ref{fig:sed} 
show that this warm component displayed large brightness variations over the last decade.
The 3.4\,{\micron} {\sl WISE} time-domain data make it clear that the disc in some periods 
was even brighter at this wavelength than at the time of the IRTF measurement. 
During these periods, the total infrared luminosity of the disc likely exceeded 
the 10\% of the stellar luminosity. Of the known EDDs, only the warm disc of 
V488\,Per exhibits a higher fractional luminosity \citep{zuckerman2012,rieke2021,sankar2021} 
and before its fading, a similarly high fractional luminosity was measured for TYC\,8241-2652-1 
\citep{melis2012}.

Assuming blackbody grains, the dust temperatures of 1040\,K and 530\,K correspond to 
radial distances of 0.09 and 0.33\,au from the star, respectively.  
Obviously, the inferred radii have a large uncertainty, since the measured emission is 
partly optically thin and partly optically thick, and the composition and size 
distribution of the dust is poorly known.

\subsection{Variability of the disc emission} \label{sec:diskvar}
At wavelengths longer than 5\,{\micron}, data are available for three epochs between 1983 and 2010, 
from {\sl IRAS} (1983), {\sl AKARI} (2006--2007), and from the cryogenic phase of {\sl WISE} (2010). 
These measurements do not suggest significant changes in the 9--20\,{\micron} range, though 
because of the sparse sampling, it cannot be ruled out that they were present in the periods 
not surveyed.  
For shorter wavelengths ($<$5\,{\micron}), there are measurements from 1999 to 2021. They
indicate the presence of a hot dust component in the disc in all studied times. In the  
earliest 2MASS near-IR observations, there is significant, 3.4\,$\sigma$ excess 
in the $K_\mathrm{s}$ band. 

\begin{figure*}
\begin{center}
\includegraphics[scale=.40,angle=0]{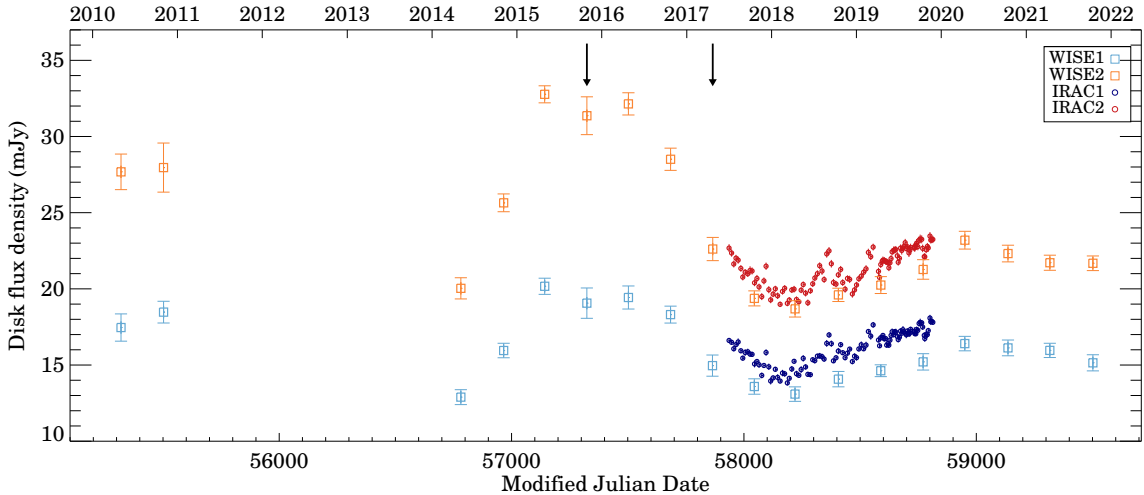}
\caption{ {Mid-infrared light curves of the disc fluxes (the measured excess emissions) of TYC\,4209 
from 2010 to 2020. The plotted semi-annual {\sl WISE} data points were derived by averaging the 
available good quality single exposure measurements in each available observing window
(Sect.~\ref{sec:wisedata}). Arrows mark those observation windows in which significant 
flux changes have been detected in the single exposure data \citep[][]{moor2021}.   
}
\label{fig:wiseiraclc}
}
\end{center}
\end{figure*}

\begin{figure*}
    \centering
    \includegraphics[width=0.45\textwidth]{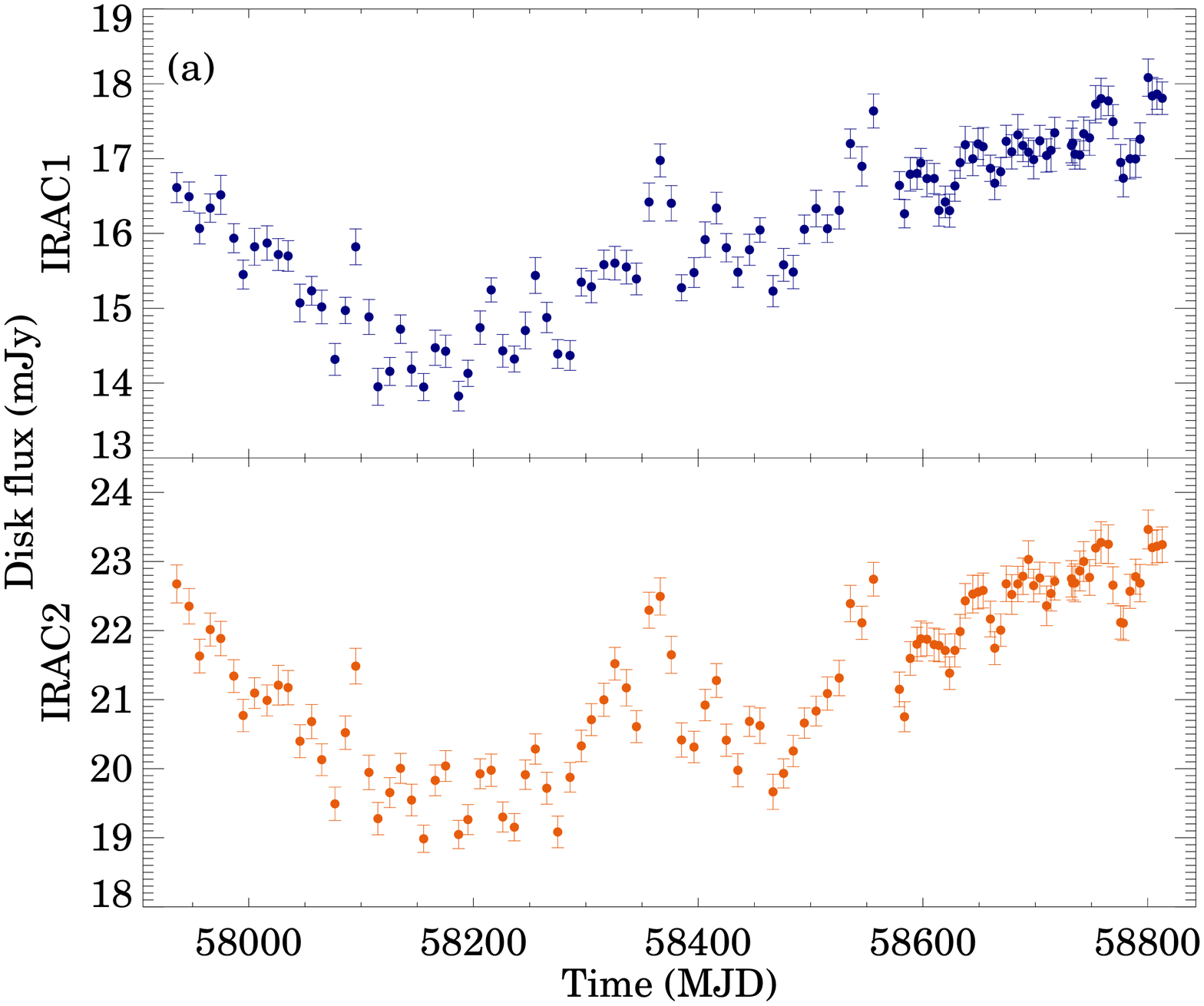}
    \includegraphics[width=0.45\textwidth]{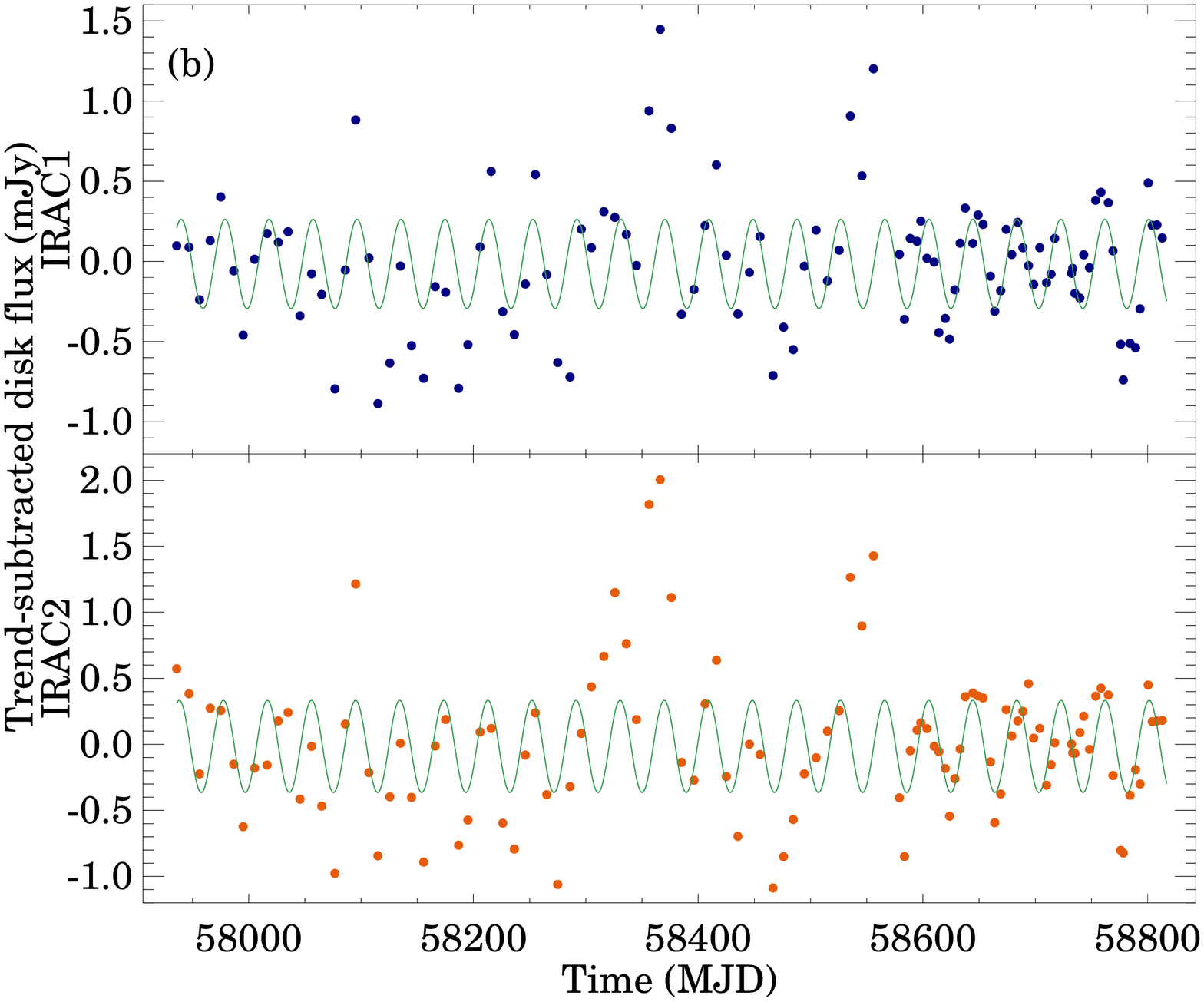}
    \includegraphics[width=0.45\textwidth]{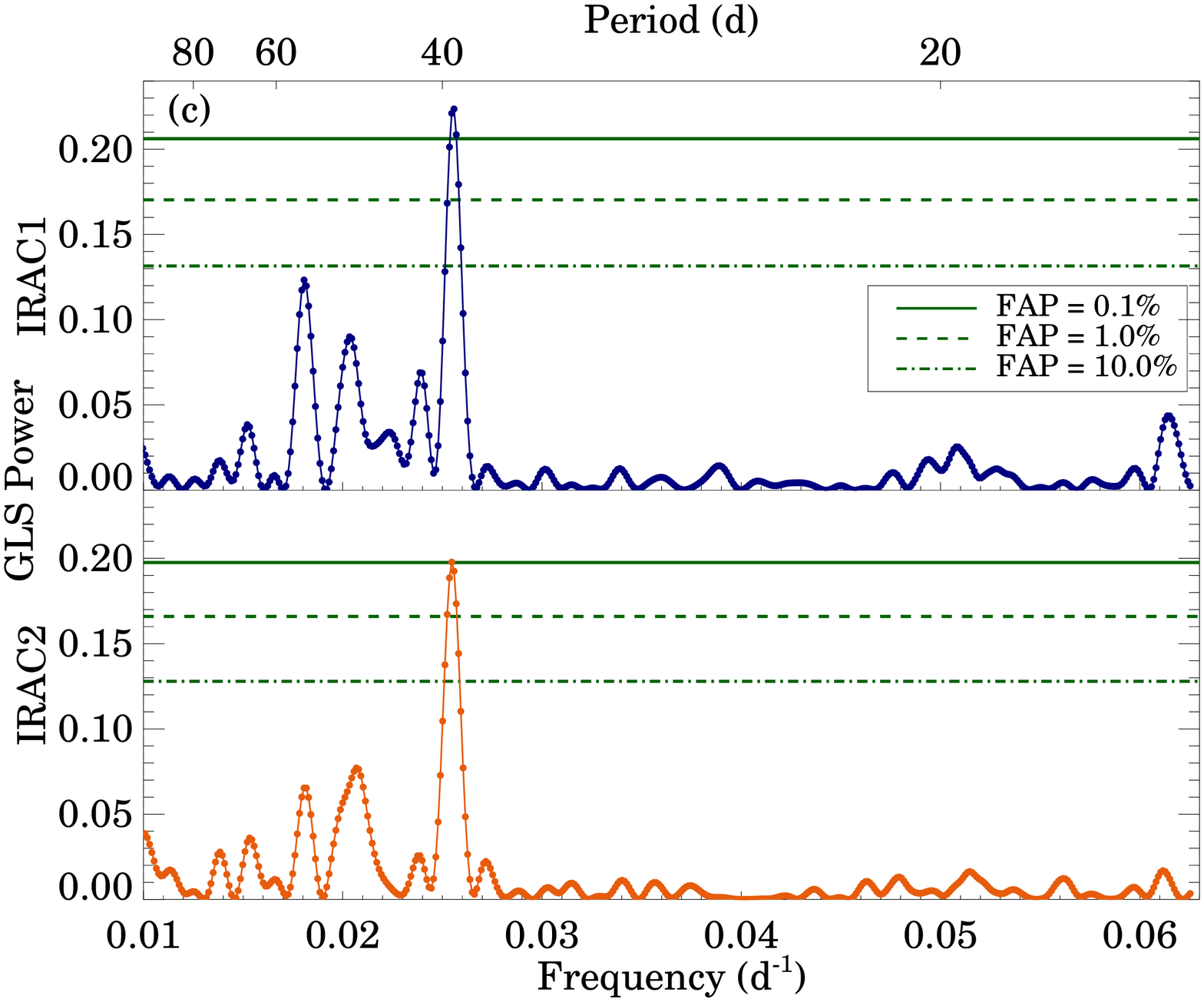}
    \includegraphics[width=0.45\textwidth]{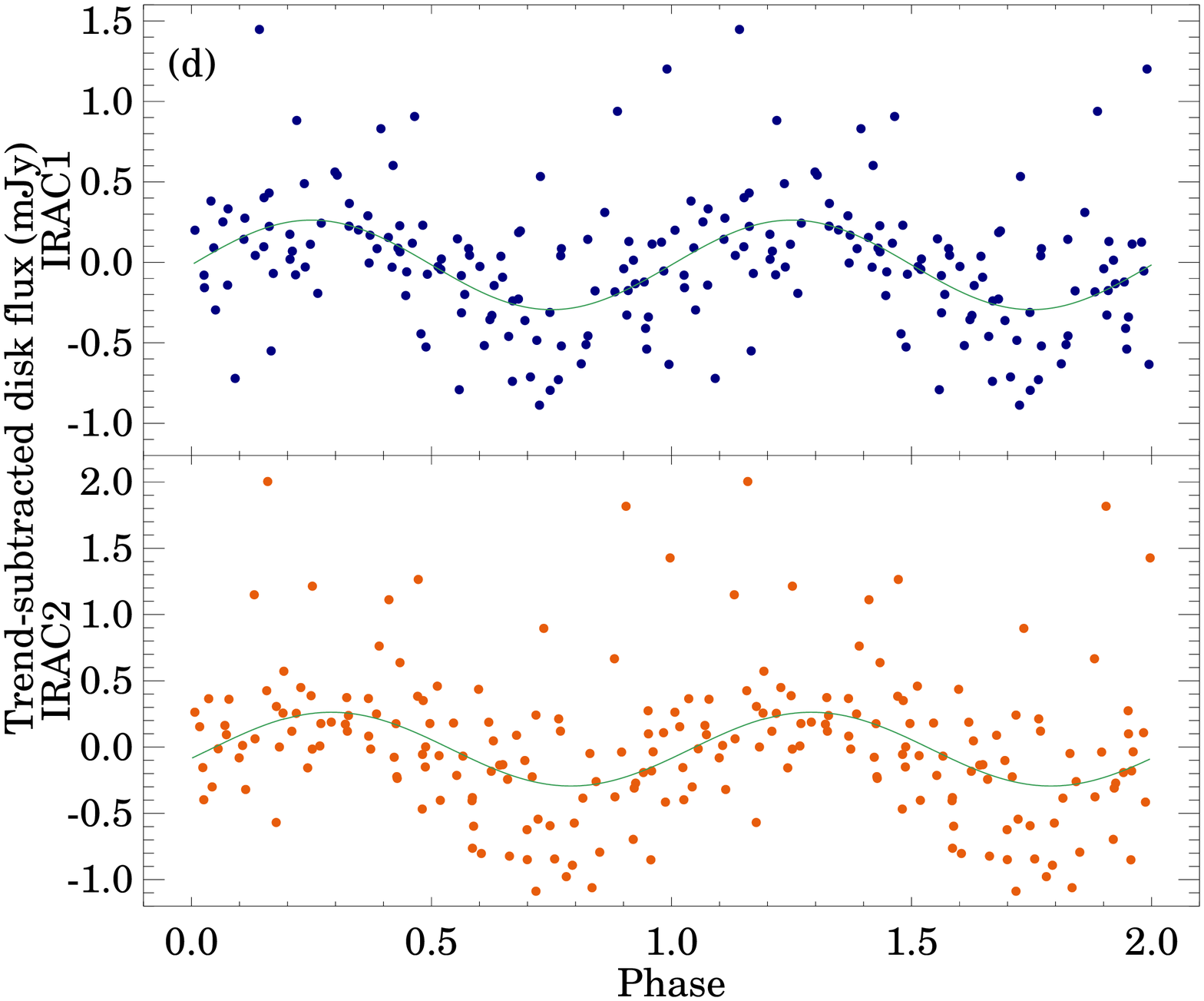}
    \caption{(a) Light curves of the measured excesses at 3.6{\micron} (IRAC\,1) and at 4.5{\micron} (IRAC\,2) 
    for the period of the {\sl Spitzer} monitoring project. (b) Light curves of the trend subtracted 
    measured excesses with the best-fit GLS models (green curves). 
    (c) Generalized Lomb--Scargle periodogram of the trend subtracted excesses. 
    (d) Phase-folded light curves of the trend subtracted excesses using the periods associated to the 
    maximum peaks found in the GLS analysis. The green curves show the best-fit GLS models.  
    }
    \label{fig:periodanalysis1}
\end{figure*}

Figure~\ref{fig:wiseiraclc} shows the time behavior of the 3--5\,{\micron} 
disc emission between 2010 and 2022. The excess data were computed by subtracting 
the corresponding photospheric flux densities from the {\sl WISE} (Sect.~\ref{sec:wisedata})
and {\sl Spitzer} IRAC measurements (Sect.~\ref{sec:spitzerdata}). 
In the common observing period, the light curves obtained with the two instruments 
-- with different time resolution and with some deviation in the flux levels due 
to the different filters -- show the same trend.
The {\sl WISE} data show changes on a longer, annual timescale between 2014 and 2022. As described in \citet{moor2021}, 
a significant brightening of the disc started sometime in 2014, followed by a roughly constant 
brightness plateau between April 2015 and April 2016, after which the disc faded, returning to 
prebrightening flux levels by early 2018. Then the flux of the disc started to increase again at 
both wavelengths, although more slowly and with lower amplitude than in 2014.
The data points after mid-2020 indicate a slow fading of the disc.
Our {\sl Spitzer} monitoring project, which run between July 2017 and November 2019, 
gives a more detailed picture about disc flux variations at 3.6\,{\micron} and 
4.5\,{\micron} in the last part of the long fading phase and in the new brightening phase 
(Fig.~\ref{fig:wiseiraclc}).
At this time, the disc exhibited a peak-to-peak flux 
variation of 26\% and 21\% at 3.6\,{\micron} and 4.5\,{\micron}, respectively. 
The light curves (Fig.~\ref{fig:periodanalysis1}a) show that rapid, relatively large amplitude variations
are superimposed onto otherwise quite smooth changes 
in the initial phase of the brightening between June 2018 (MJD$\sim$58275) and 
April 2019 (MJD$\sim$58580). 

The high precision uninterrupted {\sl Spitzer} time-domain data allow us to search 
for possible periodic modulations in the disc flux changes. For the period analysis
we employed three different methods, the Generalized Lomb--Scargle periodogram \citep[GLS,][]{zechmeister2009}, 
the phase dispersion minimization \citep[PDM,][]{stellingwerf1978}, and the Weighted Wavelet Z-transform 
\citep[WWZ,][]{foster1996}. Prior to these analyses, the general flux trend was subtracted from the 
light curves (Fig.~\ref{fig:periodanalysis1}b). For this detrending, we used a LOWESS
 algorithm with a window of 270\,days as implemented in the 
\texttt{w\={o}tan} package 
\citep{hippke2019}. The GLS periodograms (Fig.~\ref{fig:periodanalysis1}c) were calculated using the IDL version of the publicly 
available code\footnote{\url{https://github.com/mzechmeister/GLS}}. For both light curves, 
the strongest peak on the periodogram is at a frequency of  
0.0255$\pm$0.0001d$^{-1}$ that corresponds to 
a period of 39.2$\pm$0.2\,days. 
The False Alarm Probabilities (FAPs) of the peaks are 3.2$\times$10$^{-4}$ 
and 9.9$\times$10$^{-4}$ for the 3.6 and 4.5{\micron} data sets, respectively, implying that the peaks are real. 
These FAP values  
are computed using a bootstrap approach \citep{vanderplas2018}. 
The derived amplitudes of the modulation are 0.28$\pm$0.05\,mJy and 0.35$\pm$0.07\,mJy 
at 3.6 and 4.5\,{\micron}, respectively. The trend-subtracted light curves folded with the corresponding peak periods, 
together with the fitted sine curve are displayed in Figure~\ref{fig:periodanalysis1}d. 
As Figure~\ref{fig:periodanalysis1}c shows, the FAPs of the other peaks in the periodograms are 
higher than 0.1, thus they are probably not real. 
The PDM method allows to reveal periodic variations even in data showing no sine-like brightness
changes. Utilizing its implementation in \texttt{PyAstronomy}\footnote{\url{https://github.com/sczesla/PyAstronomy}} 
\citep{pya}, we found that the highest power periods for both time series are at $\sim$39.4\,days, 
which is consistent with the results of the GLS analysis. 

\begin{figure}
\centering
\includegraphics[width=0.45\textwidth]{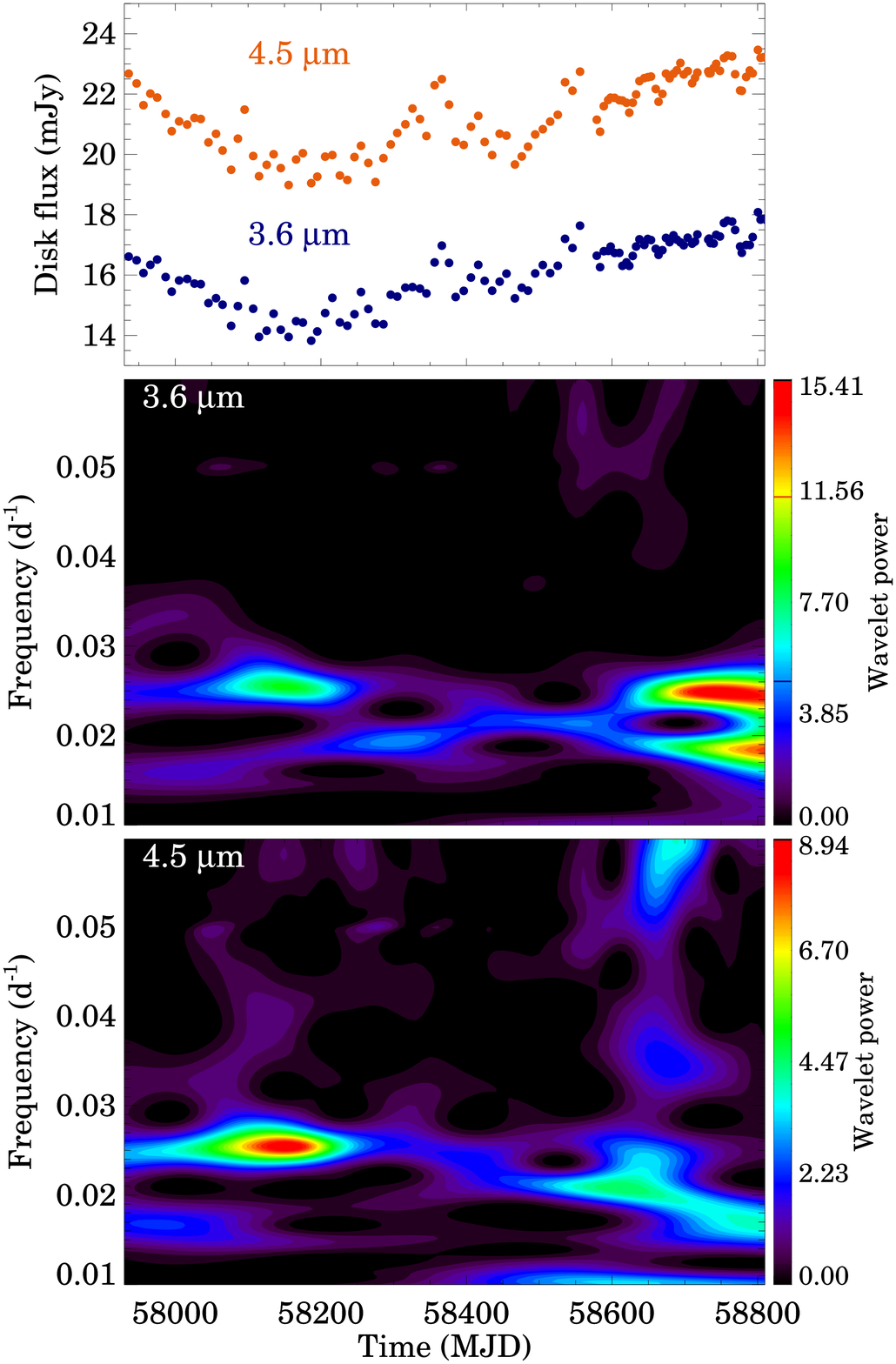}
\caption{Two dimensional contour maps of the WWZ power spectra computed for the 
    flux density time series derived for the disc at 3.6\,{\micron} (mid panel) and 4.5\,{\micron} (bottom panel). 
    For the sake of easy comparison with WWZ results, the light curves of the disc are also 
    plotted (top panel).  
    }
\label{fig:wwz}
\end{figure}

To check whether the detected period is present 
over the whole time interval under study, we applied the WWZ algorithm (using a python based implementation\footnote{\url{ http://doi.org/10.5281/zenodo.375648}}) on the trend-subtracted light curves of the disc.  
The resulting WWZ maps (Figure~\ref{fig:wwz}) show that although the frequency of $\sim$0.025\,d$^{-1}$ is present at some 
level throughout the time series examined, it is much weaker at MJDs from 58280 to 58600. 
This coincides with the onset of the brightening that is marked by larger 
amplitude changes. The data points in Figures~\ref{fig:periodanalysis1}b and \ref{fig:periodanalysis1}d 
that differ most from the GLS model also belong to this interval of time. 

Considering the results of the WWZ analysis, the GLS periodograms were calculated separately 
for the early part of the light curves (MJD: 57935--58300) when the fluxes show a downward 
trend and for the last part of the time series (MJD: 58580--58850) that are characterized by 
a relatively smooth 
upward trend. For the former, a dominant period of 39.8$\pm$0.7\,d (with a FAP of 3.3$\times 10^{-3}$) 
was found at 3.6\,{\micron} and a period of 39.5$\pm$0.6\,d at 4.5\,{\micron} (FAP=1.5$\times 10^{-4}$), 
with amplitudes of 0.37$\pm$0.07\,mJy and 0.45$\pm$0.08\,mJy, respectively.  
By examining the late part of the data, we found dominant peaks at 40.6$\pm$0.6\,d (FAP=6.1$\times 10^{-6}$) 
and 39.4$\pm$0.7\,d (FAP=5.1$\times 10^{-4}$) on the periodogram, with 
amplitudes of 0.29$\pm$0.04\,mJy and 0.32$\pm$0.05\,mJy in the 3.6 and 4.5\,{\micron} 
data, respectively. The phases obtained in the analyses of the early and late light curve 
sections are the same within the uncertainties. 

\begin{figure*}
\centering
\includegraphics[width=0.45\textwidth]{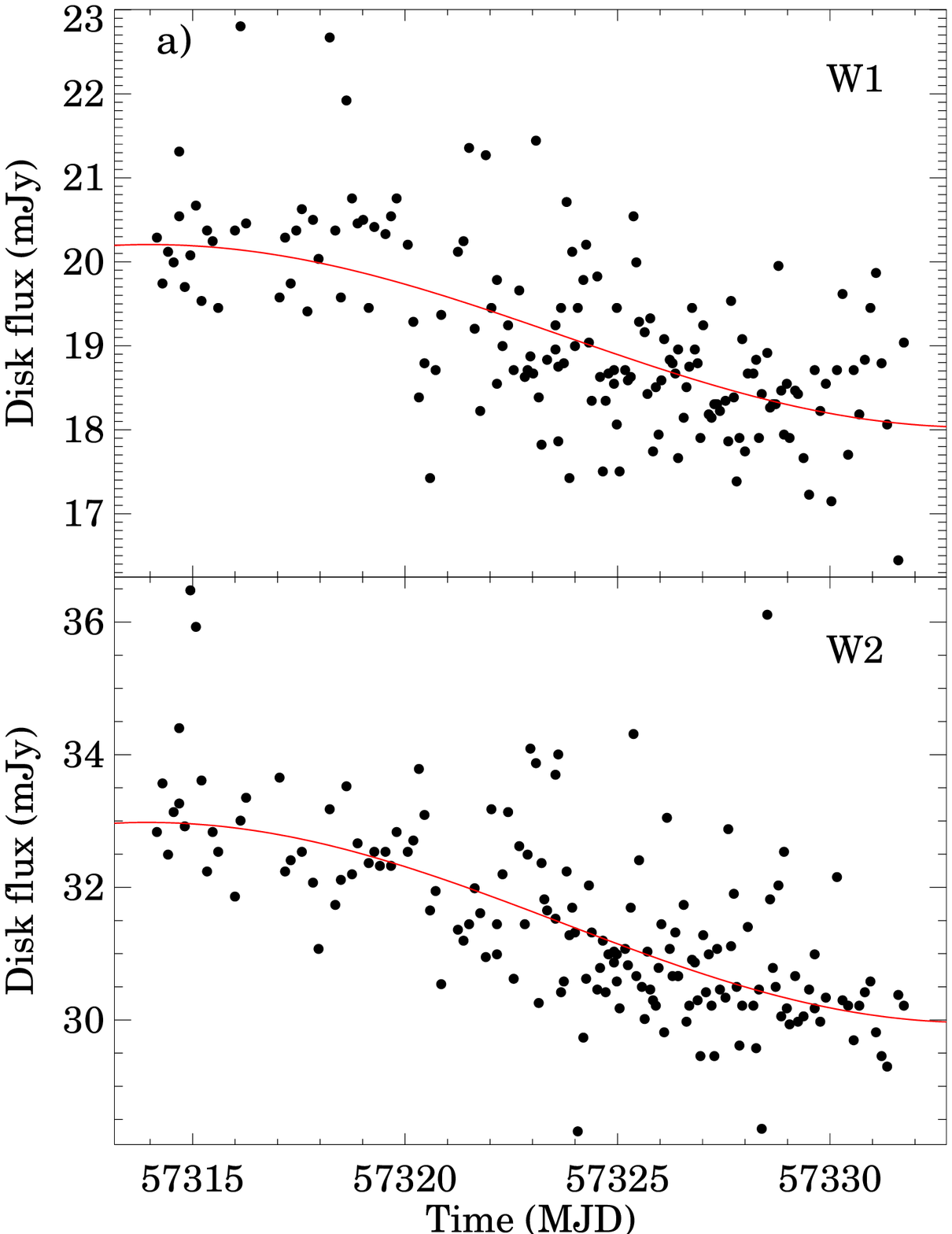}
\includegraphics[width=0.45\textwidth]{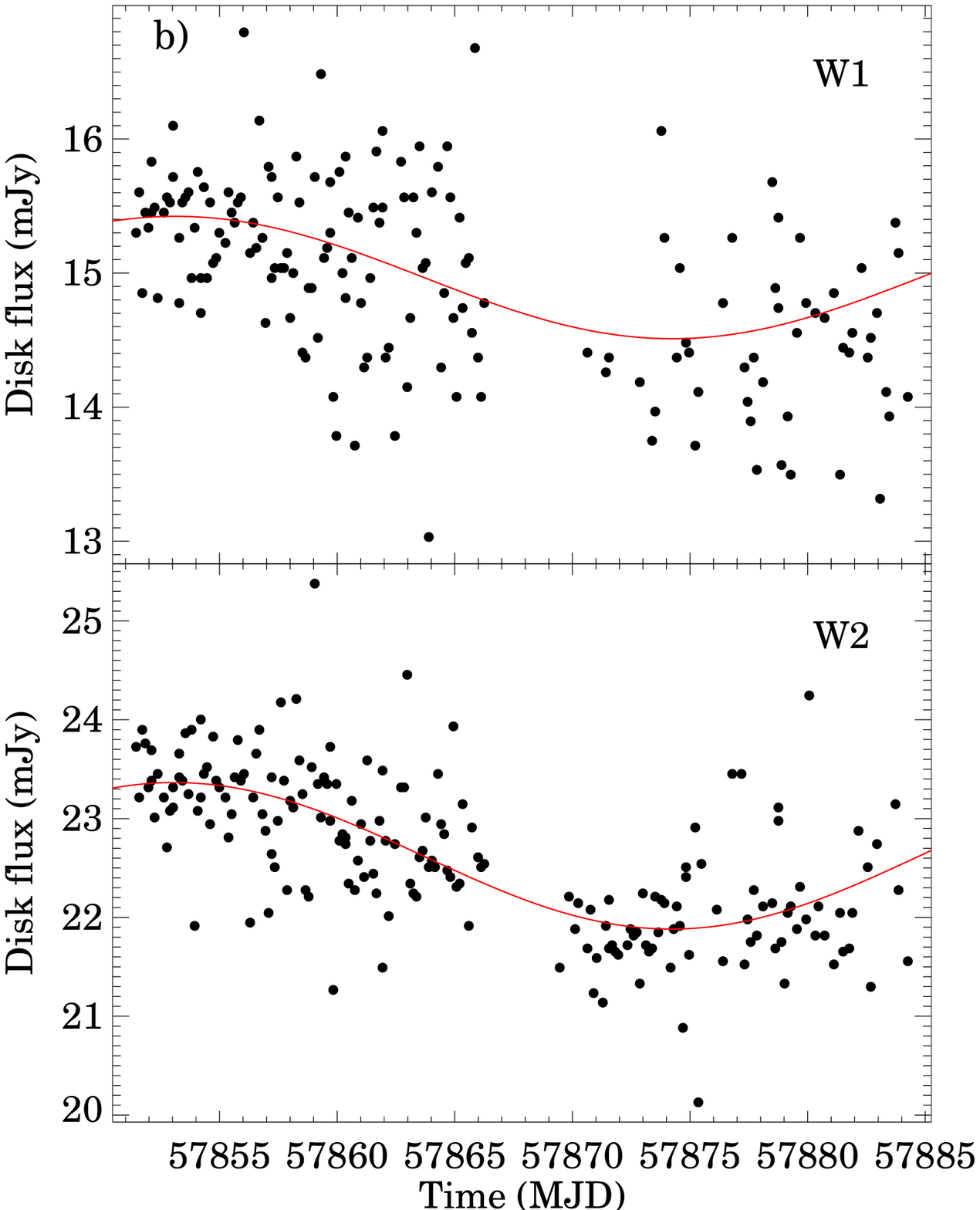}
\caption{Disc fluxes in WISE $W1$ and $W2$ bands based on single exposure photometric data  
with the best fit sine model (red line) 
(see Sect.~\ref{sec:diskvar}).  
    }
\label{fig:wise8}
\end{figure*}

To find out more about how long this short term periodic variation has been present, we explored
if there was any sign of it in the {\sl WISE} single exposure data. As we noted in Sect.~\ref{sec:wisedata}, 
these data samples are measured in observing windows with a cadence of 6~months and with their length 
between 17 and 40\,days, they are typically shorter than the period of the observed variation. Moreover, 
their typical accuracy is well below that of {\sl Spitzer} photometry. Actually, by examining each window 
separately, we found significant flux changes only in two of them \citep{moor2021}. Both sets of 
measurements preceded the {\sl Spitzer} monitoring project (Fig.~\ref{fig:wiseiraclc}).
To test whether the given {\sl WISE} $W1$ and $W2$ light curves are consistent with the periodic 
modulation found in the {\sl Spitzer} data
we fitted them with a sine function, where the phase was fixed to the value inferred from the analysis 
of the early part of the IRAC light curves (MJD: 57935--58300), while the period and amplitude 
parameters were free parameters. In the fitting, we adopted a constant background, i.e. we assumed 
that there was no major change in the longer-term trend over the given period.    
Fig.~\ref{fig:wise8} displays the {\sl WISE} light curves with the best fitted models. 
For the earlier data set, which preceded our monitoring programme by $\sim$600\,days, 
we obtain a best period of $\sim$39\,days in both bands, and amplitudes of 1.1\,mJy and 
1.5\,mJy in $W1$ and $W2$, respectively. Due to the short, only 17 days long measurement window, 
it cannot be ruled out that, despite the reasonably good fit (Fig.~\ref{fig:wise8}a), 
we are in fact only seeing a fading of the object in line with the prevailing 
long-term trend.
What contradicts this is that by the time of the next {\sl WISE} measurements, 
the disc had become brighter (Fig.~\ref{fig:wiseiraclc}) in contrast to the strong downward trend observed here.
At $\sim$33~days, the other measurement window is particularly long (Fig.~\ref{fig:wise8}b). 
For these observations we find the best fit with a period of $\sim$42\,days, 
and with amplitudes of 0.5\,mJy in the $W1$ and 0.7\,mJy in the $W2$ band, respectively.
The obtained reasonably good fit and the proximity of the measurement to the 
monitoring program -- last exposure was taken just $\sim$52 days before the 
start of the {\sl Spitzer} observations -- suggest that the $\sim$40\,days periodic modulation 
identified there may have already been present during this period.

Using the GLS algorithm, we also investigated whether the ratio of the IRAC1 and IRAC2 disc fluxes 
(i.e., the colour temperature of the disc) exhibits periodicity. We found no significant 
peaks in the periodograms of these data.

\section{Discussion} \label{sec:discussion}

\subsection{Grain dynamics} \label{sec:graindynamics}
\begin{figure}
\centering
\includegraphics[width=0.50\textwidth]{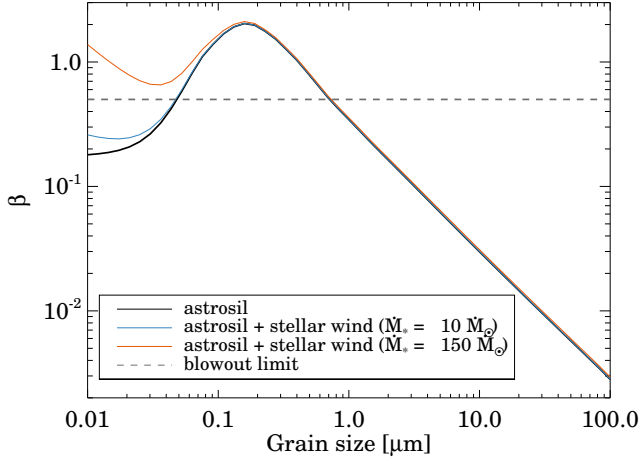}
\caption{The $\beta$ ratio as a function of grain size for astrosilicate grains.
     Influence of the stellar wind pressure force is also explored for wind 
     mass loss rates of $10 \times$ and $150 \times \dot{M}_{\rm \odot,wind}$. 
     The blowout limit of $\beta$=0.5 is shown by a dashed horizontal line. 
    }
\label{fig:beta}
\end{figure}

Interactions with the stellar radiation and wind play an important role in the
 dynamics and removal of smaller solids produced by collisions 
of larger bodies in the disc. The extent to which a grain is affected by 
the radiation and stellar wind pressure forces is characterized by the 
$\beta$ value, the ratio of these forces to the gravitational force, 
$\beta = \frac{F_{\rm rad}+F_{\rm sw}}{F_{\rm grav}}$. 
Assuming that grains are released from parent bodies on circular orbits, particles 
with $\beta > 0.5$ are blown out from the system on hyperbolic orbits.
For compact spherical grains with radius of $s$ and density of 
$\rho$, $\beta$ can be computed as
\begin{equation}
\beta = \frac{F_{\rm rad}+F_{\rm sw}}{F_{\rm grav}} = 
\frac{3 L_*  \left( Q_{\rm pr} + Q_{\rm sw} \frac{ \dot{M}_{\rm *,wind} v_{\rm wind} c}{L_*} \right)}{16 \pi G M_* c \rho s}, \label{eq:beta} 
\end{equation}
where $Q_{\rm pr}$ is the radiation pressure efficiency,
$Q_{\rm sw}$ is the  stellar  wind coupling coefficient,
$\dot{M}_{\rm *,wind}$ 
is the stellar wind mass loss rate, 
and $v_{\rm wind}$ is the stellar wind velocity \citep[e.g.][]{strubbe2006}.
The solid black curve in Fig.~\ref{fig:beta} displays $\beta$ as a function of 
grain radius for astrosilicates 
\citep{laor1993} with a density of 2.7~g~cm$^{-3}$ under the influence of 
radiation pressure alone. The radiation pressure efficiency was calculated from Mie
theory using the code developed by \citet{bohren1983}.
Based on the results, the $\beta>0.5$ condition holds only for a narrow range of grain sizes between 0.05\,{\micron} and 
0.7\,{\micron}. We note that, these are mostly rough estimates, 
since we do not know the composition of the dust and its possible porosity, 
which can seriously affect the blowout size 
\citep[see e.g.,][]{kirchschlager2013,arnold2019}. In fact, the results from \citet{arnold2019} 
show that many realistic grain compositions and structures are more amenable to blowout 
than the astronomical silicates illustrated in Figure~\ref{fig:beta}.

Young, fast-rotating early G-type stars such as TYC\,4209 typically show 
strong stellar activity and develop substantial stellar wind. By searching 
the Second ROSAT all-sky survey source catalogue \citep{boller2016}, we found a 
source, 2RXS\,J181703.5+643340, 
that is located at an angular distance of 15{\arcsec} from 
the optical position of TYC\,4209, well within the typical positional 
uncertainty of faint ROSAT sources (Fig.~\ref{fig:chart}). Using the appropriate counts-to-energy 
conversion factor quoted in \citet{kashyap2008}, we obtained an X-ray luminosity of 
2.7$\pm$1.0$\times$10$^{29}$\,ergs~s$^{-1}$, which is $\sim$120$\times$ higher than the average 
value for our Sun \citep[2.2$\pm$1.0$\times$10$^{27}$\,ergs~s$^{-1}$,][]{judge2003}. 
This implies a moderately strong coronal activity in line with the fast rotation. 
We note that the coronal activity and the X-ray flux of a star can vary strongly, and since we have one 
measurement only, it is questionable to what extent the obtained X-ray luminosity 
can be considered representative for TYC\,4209. 
To assess the possible effect of 
the stellar wind on the grains, we need to know the wind mass loss rate 
and the wind velocity (Eq.~\ref{eq:beta}) for which we have no direct
measurements. The wind mass loss rate per surface unit is proposed to vary as a 
function of X-ray flux as $ \dot{M}_{\rm \odot,wind} / R_*^2 \propto F_X^w$ 
\citep{wood2005}. \citet{ahuir2020} argued that the $w$ power-law index falls
somewhere between 0.5 and 1.0. Taking these two boundary values, 
we obtained a mass loss rate range from 14 up to 150\,$\dot{M}_{\rm \odot,wind}$ 
($\dot{M}_{\rm \odot,wind} = 2 \times 10^{-14} {M}_{\rm \odot} {\rm yr}^{-1}$ is 
the wind mass loss rate of the Sun) for TYC\,4209. As an alternative approach, 
\citet{johnstone2015} proposed that the wind mass loss rate can be estimated 
 as
$\dot{M}_{\rm *,wind}= \dot{M}_{\rm \odot,wind} \left( \frac{R_*}{R_\odot} \right)^2 
 \left( \frac{\Omega_*}{\Omega_\odot} \right)^{1.33} \left( \frac{M_*}{M_\odot} \right)^{-3.36},$ 
where $R_*$, $\Omega_*$, and $M_*$ are the radius, the angular velocity, and the mass of the 
star. This formula yields $\dot{M}_{\rm *,wind} \sim 10 \dot{M}_{\rm \odot,wind}$ for our target 
which is close to the minimum of the mass loss values derived from the X-ray flux.
By adopting a wind velocity of 400~km~s$^{-1}$, corresponding to the average wind velocity of the Sun, 
and assuming $Q_{\rm sw} = 1$, we recalculated the $\beta$ ratios by considering the 
possible effect of stellar wind. 
As Figure~\ref{fig:beta} demonstrates, by assuming a mass loss rate of 
 $\dot{M}_{\rm *,wind} = 10 \dot{M}_{\rm \odot,wind}$, the stellar wind has observable effects only 
for grains smaller than 0.05\,{\micron}, but does not lead to blowout even for those, while 
in the case of $\dot{M}_{\rm *,wind} = 150 \dot{M}_{\rm \odot,wind}$, the wind can 
significantly increase the $\beta$ of small particles ($<$0.1\,{\micron}) and can even push them out of the 
system.

The collisional lifetime of the smallest, still bound grains can be estimated as 
\begin{equation}
\label{eqn:tc}
t_\mathrm{c,bl} (\mathrm{yr}) = 0.04 (r_\mathrm{d}/\mathrm{au})^{1.5} ({M_*/{\rm M_\odot}})^{-0.5} (dr/r_\mathrm{d}) f_{\rm d}^{-1},
\end{equation}
where $r_\mathrm{d}$ and $dr$ are the disc radius and the disc radial width, respectively 
\citep{wyatt2007}. By adopting $M_* = 1.05$\,M$_\odot$ \citep{moor2021}, $r_\mathrm{d} = 0.3$\,au, $dr/r_\mathrm{d} = 0.5$, 
and $f_{\rm d} = 0.07$ (Sect.~\ref{sec:diskproperties}), this formula yields $\sim$0.05\,yr for $t_\mathrm{c,bl}$. 
This calculation supposes an axisymmetric dust distribution,  
but if the majority of the observed debris is generated in a recent giant impact, then an asymmetric 
spatial distribution is expected, 
where most collisions occur at the so-called collision-point \citep[see Sect~\ref{sec:interpretation} and][]{jackson2012}. 
The extremely high collision frequency at this point results in an enhanced collision rate averaged over the orbit of the debris, 
that is possibly even two orders of magnitude higher than derived assuming an axisymmetric distribution \citep{wyatt2016}.
Fragmented particles with $\beta>$0.5, are expelled from the system by the radiation pressure 
roughly on an orbital timescale. Assuming that they are produced at a radius of 0.3\,au,
this timescale is $\sim$0.16\,yr or $\sim$60\,days.
 
Grains are also subject to Poynting-Robertson (P-R) effect and stellar wind drag, which cause the orbiting 
grains to feel a headwind, lose angular momentum, and drift radially inward. Under the 
P-R drag, a dust grain revolving on a circular orbit at radius $r$ reaches the star on a timescale of 
$t_{\rm P-R} ({\rm yr}) = 400\frac{r({\rm au})^2}{M_*({\rm M_\odot})}\frac{1}{\beta}$ \citep{wyatt2005}. Adopting an initial orbital radius of 0.3\,au, this yields a removal timescale of $\sim$70\,yr for 
grains with $\beta=0.5$. In debris discs hosted by young magnetically active late-type stars, 
the stellar wind drag could be dominant over that of the P-R drag. 
According to \citet{plavchan2005}, the ratio of the drag times can be estimated as
$\frac{t_{\rm P-R}}{t_{\rm sw}} = \frac{Q_{\rm P-R}}{Q_{\rm sw}} \frac{\dot{M}_{\rm *,wind}c^2}{L_*}$. 
By assuming that the $\frac{Q_{\rm P-R}}{Q_{\rm sw}}$ ratio of the coupling coefficients 
is equal to 1, and using the maximum value obtained for the wind mass loss rate  
($\dot{M}_{\rm *,wind}$=$150 \dot{M}_{\rm \odot,wind}$), 
the timescale of the wind drag is $\sim$30$\times$ shorter than the P-R timescale.  
Even this timescale is longer than the collision timescale, indicating that
the dust loss is dominated by radiation pressure blowout of small particles, 
i.e., most particles are destroyed in 
collisions and expelled by the radiation pressure faster than they can migrate 
inwards due to the P-R or the stellar wind 
drag.

\subsection{Analysis of the flux changes} \label{sec:colortemperatures}

In \citet[][fig.~9]{moor2021}, we analysed the {\sl WISE} W1/W2 colour temperature evolution of the 
TYC\,4209 system. For that we converted the W1/W2 magnitudes to flux densities using the 
official zero point values of the two {\sl WISE} bands. Then we plotted the data points in a linear 
W1 (Jy) vs. W2 (Jy) diagram. Any distribution of points following a straight line would 
correspond to a variability in which the W1/W2 colour temperature is constant. The data 
presentation method somewhat resembles that of the 'flux variation gradient method' used 
to separate the contributions of an AGN from its host galaxy \citep[see e.g.][]{pozonunez2014}. 
The advantage of our method over analysing the temporal evolution of the W1--W2 magnitude 
difference is that if the observed flux is the sum of more than one emitting components 
(which is suggested in our case by Fig.~\ref{fig:sed}), 
then the W1--W2 colour would not give the colour temperature of the variable component but 
provide some weighted average of the temperatures of the different components, which is 
less straightforward to interpret.

In the case of TYC\,4209, \citet{moor2021} suggested that before MJD~58000 (mainly corresponding 
to the 2014--17 brightness maximum) the mid-IR fluxes increased then decreased with a constant colour 
temperature of $T$(W1/W2)$\sim$750\,K. We interpreted this as the appearance/disappearance of a 
warm dust population with a constant temperature. 
We also mentioned that after 2017 this warm component 
coexisted with a new hotter component that dominated the integrated flux.

\begin{figure}
\centering
\includegraphics[width=0.47\textwidth]{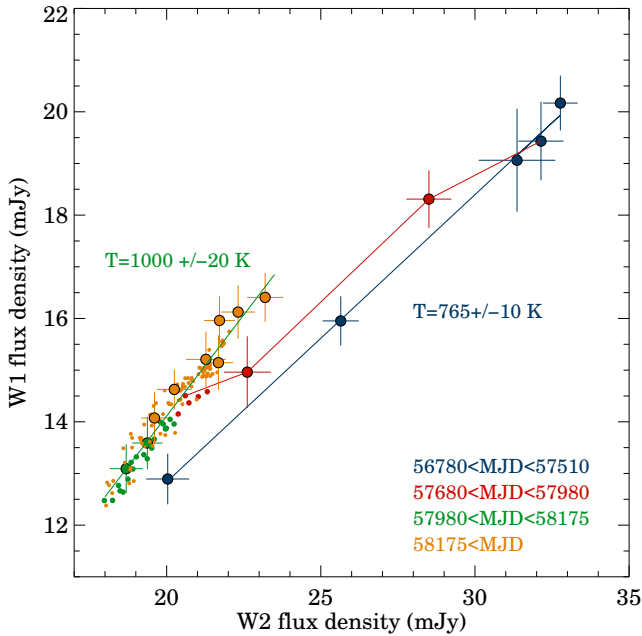}
\caption{Cross plot of 3.4 and 4.6\,{\micron} disc flux densities. 
Larger circles show the semi-annual {\sl WISE} data measured in the NEOWISE Reactivation mission phase, 
while the smaller circles denote time domain {\sl Spitzer} photometry transformed 
into the {\sl WISE} photometric system. The different colours correspond to different time intervals.}
\label{fig:colorchanges}
\end{figure}

Since this analysis, availability of four new {\sl WISE} data points (for 2020 and 2021) and 
a number of {\sl Spitzer} measurements helped to refine the above picture. For our present 
investigation, we transformed the {\sl Spitzer} fluxes into the {\sl WISE} photometric system 
by means of the following linear formulae derived by interpolating the {\sl Spitzer} light curves 
at the epochs of the {\sl WISE} measurements: $F(W1)=0.904*F(\mathrm{IRAC1})+0.811$\,mJy, $F(W2)=0.791*F(\mathrm{IRAC2})+1.446$\,mJy. 
Fig.~\ref{fig:colorchanges}, which is an updated version of fig.~9 in \citet{moor2021}, shows the flux 
evolution including all NEOWISE and {\sl Spitzer} photometric points.

In the first part of the studied period, which coincides with the rising branch and 
the peak of the 2014--2017 brightness maximum (blue circles), the W1 vs. W2 flux 
distribution follows a linear trend, whose slope corresponds to a colour temperature of 
765$\pm$10\,K. Adopting this value as the physical temperature of the radiating 
dust grains, the peak-to-peak W2 flux change can be explained by an increase of 
0.005\,au$^2$ in the total emitting area. Later, in the last part of the light curve 
from MJD=57980 until now, the data points again outline a linear locus (green and 
yellow circles), whose colour temperature, however, is significantly higher, reaching 
1000$\pm$20\,K. This temperature is in good agreement with the value found in our 
IRTF observation (Fig.~\ref{fig:irtfspectra}). In the first section of this period 
(57980$<$MJD$<$58175, green circles) the fluxes decrease until a minimum in early 
2018. Using the above high colour temperature value, this drop of flux suggests a 
change in the emitting area of $\sim$0.0004\,au$^2$. Then, the disc brightens again 
until MJD=58950, followed by a slow fading phase that is still going on 
(yellow circles). Within this whole interval (58175--now) the peak-to-peak change 
in the emitting area is about $\sim$0.0009\,au$^2$. There is also an intermediate time 
interval between MJD=57500 and 57980, covering the drop from the first peak (2016 Apr) 
until about 2017 Aug (red circles). In this time interval the disc moved from the 
765\,K locus toward the hot, 1000\,K locus. This may suggest the gradual disappearance of 
the 765\,K dust component, and the parallel building up of the 1000\,K emitting dust 
surface. Interestingly, after an initial build-up of that hot component, its emission 
dropped until early 2018, then started increasing again.

\subsection{Interpretation of the disc variability observed between 2014 and 2021} \label{sec:interpretation}

The episodic dust production leading to the EDD phenomenon is generally 
attributed to a recent giant impact \citep{melis2010,su2019}. 
In the following, we will analyse the infrared variations of the disc
in the framework of models of giant collisions \citep{jackson2012,jackson2014,kral2015,watt2021}.

\subsubsection{Variations on yearly timescale} \label{sec:yearlyvar}
As Figure~\ref{fig:wiseiraclc} shows, between 2014 May and 2015 April the disc flux increased by 
$\sim$60\% in both $W1$ and $W2$ bands. After that, it faded back to its early 2014 flux level 
over about two and a half years. 
Based on our analysis, the observed variations between 2014 and 2018 
are best explained   
by the appearance and subsequent disappearance of a cloud of fresh dust
at a temperature of about 765\,K with a minimum emitting area of 0.005\,au$^2$.
This is $\sim$2.5$\times$ larger than the emitting area estimated for the 2013 collision in ID8 \citep{su2019},
but 10$\times$ smaller than that obtained for the recent dust production in HD\,166191 \citep{su2022}. 

Energetic, high velocity collisions cause a fraction of the impact material to vaporize.
As the emerging vapour cloud cools, it rapidly condenses into small spherules having 
a narrow range of sizes that typically peaks between 0.1 and 10\,mm \citep{johnson2012}. In 
addition to vapour condensates, the debris material also contains larger boulders which are made 
of unaltered material. If the fresh dust cloud around TYC\,4209 is indeed related to a giant collision, then it 
likely consists mainly of small spherules that contribute to the 
IR emission of the disc immediately after their appearance. Right after the collision, the emerging 
debris is still localized and forms a clump that moves with the progenitor. This clump phase lasts typically no 
longer than a few orbits \citep{jackson2014,watt2021}. As the optically thick clump spreads due to Keplerian shearing, a spiral 
structure forms while the brightness of the disc increases rapidly. 
The spiral structure can then persist for up to 100 orbital periods before, becoming tighter and 
tighter, it is completely smeared out, leading to a morphology when the debris material is smooth 
but still highly asymmetric \citep{jackson2014}. 
At the same time, spherules are quickly ground down in collisions between each other 
and with the background disc material. This can further increase the emitting surface for a while, 
however, as the radiation pressure can expel blowout grains from the emerging optically thin 
regions, this population disappears and the disc brightness drops.

Assuming that the above inferred minimum emitting area is related to a dust grain population 
with sizes ($s$) ranging from radii of 0.7\,{\micron} (the blowout limit) to $s_\mathrm{max}$ 
and having a power-law size distribution with an exponent of $-$3.5 ($n(s) \propto s^{-3.5}$), 
then the mass of these grains is 
$M_\mathrm{dust} \geq 1.8\times10^{-6} (s_\mathrm{max}/1\mathrm{mm})^{0.5} (\rho/2.7\mathrm{gcm^{-3}})$\,M$_\oplus$.
The escaping vapour mass is only a fraction of the total mass of the colliding bodies ($f_\mathrm{vap} = M_\mathrm{vap} / M_\mathrm{tot}$). 
Therefore, the formation of the observed dust population requires 
the collision of two large bodies with mass equivalent of two spherical bodies
with radii of 
$\geq 78 (s_\mathrm{max} / 1\mathrm{mm})^{1/6} (1/f_\mathrm{vap})^{1/3} $\,km. 
By simulating giant impacts with various parameters utilizing a smoothed particle hydrodynamical
code, \citet{watt2021} obtained $f_\mathrm{vap}$ values lower than $\sim$0.2. 
Adopting this value and an $s_\mathrm{max}$ of 0.1\,mm 
(thus assuming a maximum size that corresponds the smallest 
typical spherule size within the range given by \citet{johnson2012}, 
see above), this formula gives radii of $\gtrsim$90\,km 
for the case of TYC\,4209.

It is worth noting that shock-induced vaporization requires high impact velocities 
that depend on the involved planetary material.  
For instance forsterite needs an impact velocity exceeding 6\,km~s$^{-1}$, while 
vaporization of quartz already begins at impact velocities below 4\,km~s$^{-1}$ \citep{davies2019}. 
In the case of self-stirring, when the dynamical excitation is linked exclusively to local planetesimals, 
the typical impact velocity is comparable to the escape velocity of the largest body situated in the region 
\citep{weidenschilling1980}.
The 4\,km~s$^{-1}$ mentioned above is similar to the escape velocity of Mercury in the solar system.
Of course, if other larger planets are present in the system, they may also contribute 
to the excitation, in which case the local bodies need not to be so large for the proper stirring. 

Analysis of the {\sl WISE} and {\sl Spitzer} light curves, as well as our IRTF 
spectrum, revealed a hotter $\sim$1000\,K dust component, which may have appeared 
at some point between MJD=57500 and 57980 when the previously mentioned $\sim$765\,K 
dust material was still present. As the colder grains are gradually removed, the 
new dust component became more and more dominant. The brightening of the disc after 
MJD$\sim$58175 is associated with this hot material. Where might this new dust 
component have been produced? Based on the higher dust temperature, this could 
belong to a collision that occurred at a smaller radius than the previous one. 
However, according to models of giant collisions, it is more likely that a new dust 
production event occur in roughly the same region as the previous one, 
especially due to the high density region at the collision 
point \citep{jackson2012,kral2015,wyatt2016}. In the latter 
scenario, the warmer temperature can be explained if the size distribution of the new 
grains are different from that of the previous component. In this case, one possible explanation for the 
formation of such a new component is that it comes from collisional erosion of larger 
boulders created in the impact in 2014 (or in an even earlier event). The timescale 
of this process depends on the density of the disc as well as the size distribution 
and the size-dependent strength of the boulders. As this population contains much 
larger bodies, the resulting collision cascade can have a longer lifetime than in 
the case of the vapour component \citep{su2019}. 

An unsolved aspect of this scenario,
however, how particles, formed during the fragmentation of boulders, can be so different 
from the previous dust population. 
Alternatively, it is possible that this hotter material, or at least 
a fraction of it, is related to dust forming in optically thick regions 
of the disc. Shielded from the stellar radiation, inside these regions very finely
ground particles can form as a result of frequent collisions. 
As the clump material evolves to become optically thin, these grains 
can contribute to disc emission \citep{su2020}. 
The fate of such tiny particles depends strongly on the weakly constrained strength of 
the stellar wind (Sect.~\ref{sec:graindynamics}). If the wind mass loss is large, small particles can be 
quickly repelled from the 
system under the combined effect of wind and radiation pressure. If the 
wind is weaker, however, it is possible that their dynamics is more controlled 
by drag forces and thus they can survive for a longer time (even up to a few decades) as 
they migrate inwards.
This scenario may 
better explain the different behaviour of the new dust.
Interestingly, for the specific time period we also observed a correlation between 
the optical and mid-IR brightness variations: when the disc was brighter in 
the mid-IR, the system became brighter and bluer at
optical wavelengths.
This correlation found to be weaker in the period when the colder 
dust component was dominant between MJD~56780 and 57680 (Sect.~\ref{sec:opticalvariability}).
If this phenomenon is related to the scattering of starlight by the 
emerging grains, as we suggested (Sect.~\ref{sec:opticalvariability}), then 
it may also indicate that this 
hot dust population is significantly different from the previous one 
in terms of size distribution and/or basic geometry.

The slow brightening of the disc starting from MJD$\sim$58175 is interrupted twice 
by periods characterized by more violent changes. The two sharp peaks, at MJD=58366 and 58556, 
are associated with rapid flux increases of 10 and 7\%, respectively, 
followed by rapid fading. These dust releases might be related to smaller 
collisions, whose optically thin dust product is removed after 
1--2 orbits under the influence of the stellar radiation pressure. 

\subsubsection{Periodic modulation} \label{subsec:periodicmodulation}
On top of the annual timescale flux changes, our {\sl Spitzer} observations 
revealed a shorter-term light curve modulation with a period of $\sim$39\,days. 
Using the full data set, our analysis yields modulation amplitudes of 0.28$\pm$0.05 and 0.35$\pm$0.07\,mJy 
at 3.6 and 4.5\,{\micron}, respectively. At the longer wavelength, the obtained amplitude 
corresponds to 1.6\% and 1.0\% of the average disc and the average total fluxes.
No variations with similar periods were found in the star's optical light curves, 
and even the amplitude of the detected rotational modulation is found to be $\sim$0.3\% 
of the stellar flux (Sect.~\ref{sec:opticalvariability}). The cause of the periodic changes detected in the mid-IR, 
therefore, could not be the star. 
By fitting the amplitudes with a blackbody we obtained a temperature of 930$\pm$290\,K.
With this temperature, at the distance of TYC\,4209, the observed amplitude at 4.5\,{\micron} can be reproduced by an 
emitting region having a size of $\sim$7.5$\times$10$^{-5}$\,au$^2$, which is roughly equivalent 
to the size of the Solar disc. Such a large change over such a short period of time 
can obviously be related to the circumstellar material only. 

Actually, the specific disc geometry after a giant collision can offer a natural explanation 
for the observed flux modulation. The orbits of the newly formed particles must pass through the point where the impact 
happened, thereby causing a substantial density enhancement at this so-called collision 
point. Since orbits of debris must also move through the orbital plane of the progenitor, 
another dense region, the anticollision line, is also formed
on the opposite side of the star from the collision point \citep{jackson2012}. 
At these special locations, the 
debris passes through narrower regions, and due to the high optical thickness, this can 
lead to periodic flux variations \citep{jackson2014,su2019} whose amplitude and consequently
detectability depends on the distribution of the grains' orbital parameters. 
Observable oscillations occur only for sufficiently narrow distributions of semimajor 
axes, eccentricities, and inclinations \citep{watt2021}.  
The enhancement of the density contrast at the special regions may be different, 
with the collision point becoming more dominant as the $\sigma_v/v_\mathrm{Kep}$ 
ratio increases \citep{jackson2014}.  
In a disc with edge-on or nearly edge-on geometry, 
an optically thick impact debris cloud, that becomes more elongated over time,
exhibits variable emitting cross sections as it orbits. This can also cause 
flux modulations \citep{meng2014,su2019}.  
Unless the collision occured halfway between the disc ansae or at the disc ansa, such a configuration 
results in multiple periodicity in the light curves. In the case of TYC\,4209, however, we  
see only a single period in the data. 
Moreover, the lack of detectable eclipses in the optical light curves suggests 
that the disc is probably not perfectly edge-on.  
Therefore, in the following we adopt a scenario in which we consider only the possible oscillation 
associated with the collision/anti-collision points. 
Depending on the density contrasts of these specific regions, the detected single 
period may be half of the true orbital period, or corresponds exactly to it 
(if only the oscillation related to the collision point can be detected). 
While a 39-day orbital period indicates a collision at 0.23\,au, an 78-day orbital period 
would correspond to a radius of 0.36\,au. None of these radii are in contradiction
with the rough estimates on the disc location inferred from the SED analysis (Sect.~\ref{sec:diskproperties}).

P\,1121 and ID\,8, two EDDs which were targets of five years long monitoring with {\sl Spitzer}, 
also display periodic or quasi-periodic variabilities in the mid-IR \citep{meng2014,su2019}.
For P\,1121, a $\sim$17-day period of variation with an amplitude of 0.08\,mJy was observed over the 
entire studied period. At the distance of TYC\,4209, this amplitude would be 0.21\,mJy.
In the case of ID\,8, periodic modulation was seen only for some parts of the studied time interval, 
with two intermixed periods (26 and 33\,days) in 2013 and with a single 10-day period in 
2014/2015 \citep{meng2014,su2019}. For the 10 and 26\,days periods an amplitude of 0.08\,mJy 
was derived, while the longest period was associated with an amplitude of 0.16\,mJy. 
Scaled to the distance of TYC\,4209, these amplitudes would be 0.14\,mJy and 0.28\,mJy.
So, the periods and amplitudes of the flux oscillations detected in the other two EDDs  
investigated in more detail are quite similar to those seen at TYC\,4209. 

The periodic modulation is detected mostly based on the high precision {\sl Spitzer} measurements, 
although in two previous {\sl WISE} observing windows, data hint that this $\sim$39\,days period oscillation 
was already present earlier (Sect.~\ref{sec:diskvar}). According to models, the special geometry associated 
with the putative collision should already develop shortly after the event, 
in the spiral phase of the dynamical evolution. Therefore, the appearance of flux modulation 
even at the time of the earlier {\sl WISE} observing window would not be surprising. The time 
between the first measurement of that {\sl WISE} window and the last IRAC observation is $\sim$
1500\,days, which is equivalent to $\sim$38 or $\sim$19 orbits for orbital periods of 39 or 78\,days.
By looking at the amplitudes of the oscillation over the two {\sl WISE} measurement periods and at the 
beginning and end of the {\sl Spitzer} monitoring program, we see that they tend to decrease with time 
(Sect.~\ref{sec:diskvar}). Although this finding is in line with theoretical expectations \citep{watt2021}, it 
should be treated with some caution since the {\sl WISE} amplitude results are less reliable due to 
the short observation windows. 
As our Figures~\ref{fig:periodanalysis1} and \ref{fig:wwz} show, the oscillation can 
be detected even around the apparent minimum of the disc 
brightness at MJD$\sim$58200. In this respect, the light curve of TYC\,4209 is similar to that 
of P\,1121, where the modulation was also observed at the flux minimum \citep{su2019}.
As the flux of TYC\,4209 increases again, the modulation signal first weakens and then becomes more detectable 
again. The upward trend seen in this last part of the {\sl Spitzer} light curve is predominantly 
attributed to the appearance of a hot dust component (see above). 
One possible scenario is that this hot material has appeared closer to the star than the 
previously released debris. However, the periodogram for this time interval still shows 
only one significant peak with a period of $\sim$39\,days (Sect.~\ref{sec:diskvar}). 
The lack of shorter-period modulation suggests that even if this hot dust is located at smaller radii, 
it is unable to produce detectable oscillations, for example because it is mostly optically thin.

\subsection{The long term history of the disc} \label{sec:longtermhist}

In 2010, TYC\,4209 showed elevated flux levels in $W1$ and $W2$ bands 
compared to the ones observed in early 2014 (Fig.~\ref{fig:wiseiraclc}). This suggests that an 
additional major collision event may have occurred during or right before 
this period, whose effect was significantly reduced by 2014. 
The observed decrease in flux can be explained by the disappearance of 
$\sim$870\,K dust with a minimum surface area of 0.0022\,au$^2$. 
Assuming that the radiation came from particles with size between 0.7\,$\micron$ and 
$s_\mathrm{max}$ with a
size distribution $n(s) \propto s^{-3.5}$, it corresponds to the fragmentation of two bodies 
with radii of $\geq 59 (s_\mathrm{max} / 1\mathrm{mm})^{1/6} (1/f_\mathrm{vap})^{1/3}$\,km. 
Here we suppose that we were witnessing the 
early evolutionary phase after an impact event when the vapour condensates 
dominate the emitting area. If we saw larger boulders being eroded, it would 
of course indicate a much larger total mass and larger fragmented bodies. In addition to this, the flux 
density measured by {\sl IRAS} at 12\,{\micron} is in good agreement with the 
{\sl WISE} 12\,{\micron} flux, demonstrating that TYC\,4209 possessed
similar quantity of dust already four decades ago (Fig.~\ref{fig:sed}). In summary, TYC\,4209 
has been very dusty for at least 40 years and in that time there have been at 
least two large collisional events around 2010 and 2014.

Assuming a disc comprised of solids with radii ranging from 
$s_\mathrm{min}$ to $s_\mathrm{max}$ with a size distribution of $n(s) \propto 
s^{-3.5}$, the total disc mass can be 
approximated as \citep[][eq.~15]{wyatt2008}: 
\begin{equation}
\label{eqn:mtotal}
\frac{M_\mathrm{tot}}{M_\oplus} = 2 f_\mathrm{d} {\left(\frac{r_\mathrm{d}}{\mathrm{au}}\right)}^2 \sqrt{\frac{s_\mathrm{min}}{\mathrm{\mu m}} 
\frac{s_\mathrm{max}}{\mathrm{km}}} / 0.37.
\end{equation}
Using our $f_\mathrm{d} = 0.07$ and $r_\mathrm{d} = 0.3$\,au estimates from the fitting of the 
WISE data from 2010 (Sect.~\ref{sec:diskproperties}), 
and setting the blowout grain size of 0.7\,{\micron} for $s_\mathrm{min}$ and 90\,km
for $s_\mathrm{max}$ (the estimated minimum size of the bodies assumed to be involved 
in the collision in 2014), this yields a very high disc mass of $\sim$0.3\,$M_\oplus$.
However, this estimate is hampered by significant uncertainties. 
Our fractional luminosity estimate ($f_\mathrm{d}$) refers to a period when the 
disc was brighter -- at least at 3--5\,{\micron} -- possibly due to vapour 
condensates comprising of small particles only. This overestimation of 
$f_\mathrm{d}$ may be counterbalanced by the fact that the disc may be partially optically thick.
For the smallest bodies, we set the blowout dust size derived for astrosilicate grains 
(Sect.~\ref{sec:graindynamics}). Because of the unknown composition and density, this has 
its uncertainties. Moreover, in the outlined scenario, 
the majority of collisions occur at the collision point in a presumably optically thick 
environment, where the rapid fragmentation may result in the formation of very small particles 
not repelled by the radiation pressure, but are mostly removed by the drag 
forces on a longer timescale. The potentially most serious uncertainty relates to the size 
distribution of bodies. Shortly after a giant collision the size distribution of the 
boulder population can differ severely from the adopted one, expected for a 
steady-state collisional cascade involving self-similar bodies 
\citep{wyatt2016}. Thus we need to keep in mind that a steeper size distribution can give
orders of magnitudes lower total mass than the above one.

Another possible approach to constrain the amount of solids involved in collisions is to estimate the mass-loss 
from the available data. 
Considering the mass of particles just above the minimum grain size in the collisional 
cascade, $M_\mathrm{s_{min}} = 5\times10^{-9} \rho s_\mathrm{min} 4 \pi r_\mathrm{d}^2 f_\mathrm{d}$ 
\citep[][]{matra2017}, and the collisional timescale of blowout grains (Eq.~\ref{eqn:tc}), 
we obtain an $\dot{M}_\mathrm{loss}$ 
of $\sim$1.6$\times$10$^{-5}$\,M$_\oplus$ yr$^{-1}$ and thus a mass loss of 
$\sim$6.5$\times$10$^{-4}$\,M$_\oplus$, i.e. $\sim$4$\times$ of the mass of Ceres, over the past 40 years. 
Since because of the special influence of the collision point, the collisional timescale is likely 
much shorter than what can be derived from Eq.~\ref{eqn:tc} (Sect.~\ref{sec:graindynamics}), 
and the 40-year lifetime is only a lower limit, the total mass loss was probably much higher. 
Given that the disc is still present, this also implies that the total quantity of debris 
generated in putative early giant impact must have been significantly larger. 

Considering the quite large disc mass estimates and that the analysis of 
the mid-IR light curves of the last 10 years indicate at least two major collisions,  
it is reasonable to assume that what we have seen in the last 4 decades at TYC\,4209 is the 
result of a previous collision involving larger planet-sized bodies. 
This has resulted in a very large number of fragments, including hundreds kilometer sized bodies, 
whose subsequent collisions with each other or with the progenitor -- likely in the vicinity 
of the collision point of the original major event -- produce newer and newer debris clouds.
If this is the case, the orbital radius derived from our study based on the collision in 2014 may give 
an idea about the approximate location of the original large collision and, if the progenitor was 
not destroyed, about its position. Comparing the long term variability of TYC\,4209 with that of 
ID\,8 and P\,1121, it clearly resembles more the former one whose time-domain data also show evidence 
of two violent impacts happened with a time difference of a few years only \citep{su2019}. 
This latter object is also suspected to originate from a previous large scale embryo-embryo 
collision \citep{meng2014}.

\subsection{Origin of the disc}

In the following, we discuss processes that can lead to giant collisions in the terrestrial region and 
examine how consistent their expected activity stage is with the
age of the TYC\,4209 system. Based on the lithium content, rotation, and kinematic properties of TYC\,4209\,A, 
the age of the star was estimated to be 275$\pm$50\,Myr (Sect.~\ref{sec:intro}). For a further independent 
age dating of the system, we compared the position of the M-type companion star on a Gaia-based 
colour-magnitude 
diagram  ($M_\mathrm{G}$ vs. $G - G_\mathrm{RP}$) with those of M-type members of different well-dated stellar 
clusters, using the plot compiled by \citet[][figure~8]{popinchalk2021}. We found that the location of TYC\,4209\,B 
($M_\mathrm{G} = 11.174\pm0.007$, $G - G_\mathrm{RP} = 1.29\pm0.01$) is consistent with those of 
members of clusters with ages between 100 and 750\,Myr (Pleiades, M34, M50, Praesepe, Hyades) 
but is clearly different 
from the locus of younger groups,
providing a further argument that the system is not younger than 100\,Myr.

\subsubsection{Giant collision related to terrestrial planet formation}
Most planet formation models argue that after the dispersal of primordial gas
giant collisions among planetary embryos emerging from the protoplanetary disc 
play an important role in the final assembly of terrestrial planets \citep[][and references therein]{lammer2021}. 
These events are accompanied by the production of copious amount of debris 
material providing a natural explanation for the EDD phenomenon.
Analysis of {\sl Kepler} mission data indicates that 16.5\% of FGK-type 
main-sequence stars host at least one Earth-like planet ($R_\mathrm{pl} = 0.8-1.25$\,$R_\oplus$) 
with orbital periods up to 85\,days and 20.3\% of them have at least one 
super-Earth ($R_\mathrm{pl} = 1.25-2.0$\,$R_\oplus$) in the same orbital period range \citep{fressin2013}. 
This shows that stars similar to TYC\,4209 commonly host such planets in the region 
where the collision in TYC\,4209 may have occurred.

Simulations of the giant impact phase show that the vast majority 
of collisions occur in the first few 10\,Myr after the 
gas-rich disc dissipates, and even the last events in the series of 
collisions typically happen within 200\,Myr \citep{genda2015,quintana2016}.
Traces of such giant impacts can be found on most terrestrial planets in our 
Solar system, the most iconic of which is thought to lead to the 
formation of the Earth-Moon system \citep{wyatt2016}. 
Estimates for the epoch of the latter event -- which is thought to be one of the 
last giant collisions between planetary bodies in the Solar system -- range 
between 50 and 200\,Myr after the condensation of 
calcium-aluminum rich inclusions \citep{lock2020}, more likely occurring 
within 80\,Myr \citep{woo2022}. 
The migration models of the formation of super-Earths predict resulting 
configuration with planets in resonant chains.
As simulations show, after the dissipation of the gas disc, these 
chains tend to become dynamically unstable due to 
gravitational interactions leading to orbital crossing and giant impacts 
\citep{izidoro2021,estevez2021}. While the last collisions  
typically happen within 100\,Myr, in some cases they can be postponed until 300\,Myr 
or presumably even later \citep[300\,Myr was the limit of the simulations in][]{izidoro2021}. 
In their recent paper,
\citet{melis2021} proposed such a late-stage instability 
as the possible explanation for the formation of the EDD seen in 
the 600\,Myr old TYC\,8830-410-1 system.

Thus, the disc observed at TYC\,4209 could be the result of a collision 
related to the formation of Earth-like or super-Earth planets, but if 
so then it would be an unusually late event in light of current model predictions. 

\subsubsection{Possible role of internal dynamical instabilities}
In systems where giant planets are also present, their possible
dynamical instabilities can have a substantial effect on the 
structure of a planetary system including the formation and evolution of 
inner rocky planets. A late instability may even change the orbits of 
rocky planets that have already become stable, causing intersecting orbits and 
thereby renewing the era of giant collisions. 
Simulations suggest, however, that such kind of instabilities are more common 
during the early evolution of the system, happening typically in the first 
10\,Myr \citep{bonsor2013}.
Several properties of the Solar system suggests that a dynamical instability of 
outer planets may have had a major role in forming the currently observable configuration of 
the planetary system \citep[e.g.,][]{gomes2005,brasser2009}. When this instability 
occurred is a matter of debate. While previously, linking to the hypothetical 
Late Heavy Bombardment event, it was assumed to happen at an age of 
$\sim$700\,Myr \citep{gomes2005}, more recent results suggest 
an early instability at $<$100\,Myr \citep[e.g.,][]{morbidelli2018,mojzsis2019}.
In the latter case, the instability occurred at the same time as the formation of the 
rocky planets and may have had an effect on that process 
\citep[e.g.,][]{clement2018,desouza2021}.

\subsubsection{Possible role of external dynamical instabilities}
Dynamic instability is not only caused by internal effects.
By studying known EDD systems, \citet{moor2021} found that very wide-orbit 
pairs (with separations $>$300\,au) are significantly more common in old EDD 
systems than in the normal stellar population. This excess of wide pairs in 
old EDDs suggests that the companions may have a role in the disc formation. 
A highly eccentric companion has the potential to trigger a dynamical instability 
in the planetary system during its pericentre passage.
Alternatively, a companion with a properly large inclination with 
respect to the orbital plane of a planetary system can also result in its 
significant rearrangement by causing oscillations in the eccentricity and 
inclination of the planetary bodies via the Kozai-Lidov (KL) mechanism.
Both mechanisms can cause collisions between inner planets and a larger body 
scattered inwards from the outer regions. But the rearrangement that occurs as a 
result of instability can also be at a level that leads directly to collisions 
between planets with previously stable orbits in the terrestrial zone.

TYC\,4209 has a co-moving, co-distant companion at a projected separation of 
$\sim$6000\,au \citep{elbadry2021,moor2021}, which is an M3.5-type dwarf 
star with a mass of $\sim$0.27\,M$_\odot$ \citep{moor2021}.
Because of their long orbital period, orbital parameters of 
such wide binaries are difficult to constrain. 
Recent works, based on Gaia astrometric data, found that 
eccentricities of 
stellar pairs with separation of $>$1000\,au show a "superthermal" 
distribution, i.e. they display an excess at high eccentricity values with respect 
to the "thermal" distribution where $f(e)~de=2e~de$ \citep{tokovinin2020,hwang2021}.
So this population, to which TYC\,4209 belongs, is dominantly highly eccentric.
Actually, using astrometric measurements from the Gaia\,EDR3 catalogue, \citet{hwang2021} 
derived an estimate of $e=0.9$ for the orbit of TYC\,4209\,B with 68\% lower and upper 
limits of 0.58 and 1, respectively. Although the nonsignificant proper motion difference 
of the pair ($\mu / \Delta \mu < 3$) urges caution about the outcome 
\citep{hwang2021}, this hints at a quite eccentric orbit for the target. 

Even if we assume that the orbital inclination of TYC\,4209\,B exceeds the 
critical value with respect to the inclination of a specific third body, the KL mechanism still 
needs time to have a significant influence at the location of this body. 
The characteristic KL timescale is 
$t_\mathrm{KL} \sim \frac{m_\mathrm{tot}}{m_\mathrm{B}} \frac{P_\mathrm{B}^2}{P_\mathrm{C}} (1 - e_\mathrm{B}^2)^{3/2},$ 
where $m_\mathrm{tot}$ is the total mass of the system ($m_{\rm tot} = m_\mathrm{A} + m_\mathrm{B} + m_\mathrm{C}$),
$P_\mathrm{C}$ is the orbital period of the third body, while 
$m_\mathrm{B}$, $P_\mathrm{B}$, and $e_\mathrm{B}$ are the mass, orbital period, and 
eccentricity of TYC\,4209\,B, respectively.
By assuming a circular orbit for the companion ($e_\mathrm{B} = 0$), \citet{moor2021} 
concluded that within the system's lifetime it can excite a high eccentricity outward of 
$\sim$140\,au (considering that the first eccentricity peak is reached at 
 $t_\mathrm{KL}/2$), i.e. at most it can trigger changes in an extensive outer planetesimal belt, 
if such a belt exists. In the case of a highly eccentric companion, however, the KL timescale 
may be shortened and the effects may occur further inwards. For example, by adopting 
an eccentricity of 0.9 and a semi-major axis of 6010\,au (correspondingly to 
the projected separation) for the orbit of TYC\,4209\,B, then even a body 
orbiting at $\sim$25\,au could be affected substantially by the KL mechanism. 
This could trigger an instability affecting the possible giant planets, 
eventually causing orbit changes of other inner bodies, e.g. rocky planets \citep[e.g.,][]{carrera2016}. 

Formation models of very wide binaries predict initial orbits with high eccentricities 
\citep[][and references therein]{hwang2021}.     
Because of their large separations, such systems are weakly bound, and therefore, their 
orbits are strongly perturbed by the Galactic tide as well as impulses from other passing stars.
These effects can increase the eccentricity further, leading to 
close stellar passages between the companions from time to time \citep{kaib2013}. 
Such passages have the potential to trigger a direct dynamical instability in the planetary 
system eventually leading to collisions in the terrestrial zone.
Although the typical timescales of these external perturbation effects are of the order of Gyrs, 
in some cases it could be only a few 100\,Myr \citep{kaib2013}.
So this scenario cannot be excluded for TYC\,4209 either.

While all the scenarios outlined above may be at some level feasible for the disc formation, 
TYC\,4209 is still a somewhat odd case because of its age. If the disc is associated with 
the formation of rocky planets or some internal dynamical instability, it represents a 
very late phase within this type of models. If we see the result of an instability triggered 
by some external effect, then -- given the characteristics of those scenarios -- 
we can classify it as a relatively early event.

\section{Summary} \label{sec:summary}
Between 2017 and 2019, using the {\sl Spitzer Space Telescope} we monitored TYC\,4209, 
a system hosting a recently discovered extreme debris disc \citep{moor2021}, over 
877\,days at 3.6 and 4.5\,{\micron}. Combining these observations with other time-domain 
mid-IR data, we explored the disc variability on different timescales. 
The earliest available data, obtained by the {\sl IRAS} satellite, show that the disc was already 
extraordinarily bright 40 years ago. Based on {\sl WISE} light curves at 3.4 and 4.6\,{\micron}, 
the disc exhibited substantial changes between 2010 and 2021, the most significant of which was 
the brightening and fading between 2014 and 2018 during which its brightness changed by $\sim$60\% 
in both bands. The {\sl Spitzer} data stream gives a more accurate picture of the fading phase 
and the subsequent newer brightening of the disc after 2018, allowing to examine the mid-IR
flux change on a timescale of a week. Period analysis of these data 
revealed a significant oscillation with a period of $\sim$39\,days and amplitudes of
0.28 and 0.35\,mJy at 3.6 and 4.5\,{\micron}, respectively. 

To interpret the observed variations between 2010 and 2021, we adopted the model of 
giant collisions. We argue that, there were at least two major collision events in
the system during this period. The elevated flux levels measured in 2010 imply enhanced 
dust content which we attribute to a collision that occurred at that time or shortly 
before and whose effects largely disappeared by four years later. 2014 saw another 
big dust production event. The subsequent rapid increase in mid-IR flux can be 
explained by the appearance, and then dynamical and collisional evolution of a copious 
amount of small spherules, which were formed via quick condensation of vaporized material 
ejected in the impact. Assuming that these hot ($\sim$760\,K) grains formed in a collision of two 
identical spherical bodies with $\lesssim$20\% of their material vaporized, 
we obtained a minimum radius of $\sim$90\,km for these bodies.
This dust component gradually disappeared by the end of 2017.

Our analysis also revealed an additional even hotter 
($\sim$1000\,K) dust component that already appeared when the previously 
mentioned colder dust was still present and became dominant after its disappearance.
The brightening of the disc after 2018 can be mainly attributed to this hotter debris. 
It is conceivable that this material may also be related to the aforementioned 
giant impact, either being made of collisional fragments of larger boulders formed 
in the impact or very fine dust produced in optically thick regions of the debris cloud. 
This scenario assumes material emerging at a similar radial distance to the previous 
dust population. The significantly higher temperature of this new component then 
would indicate grains with very different size distribution from the previous ones.
It cannot be ruled out, however, that the higher dust 
temperature indicates a dust cloud appearing closer to the star instead.

According to models of giant impact, the periodic modulation of the disc flux 
might be related to the special geometry of the collisional point and the part of the 
dust ring opposite it (anticollision point). Assuming a circular orbit, depending on the 
density contrast of these specific regions, the measured 39-day period
can imply a dust orbital radius of 0.23\,au if the oscillation is associated with 
the collisional point only, or 0.36\,au if both regions have a significant effect 
on the light curves. 

The fact that the disc was already peculiarly dust rich 40 years ago, suggests 
that the dust production events observed after 2010 may be the aftermath of an earlier 
more catastrophic collision perhaps involving a planet-sized body.  
The continuous collisional erosion of different sized bodies ejected in this event 
might ensure the sustenance of dust, while sporadic collisions between the largest fragments 
can increase the dust content of the disc significantly from time to time.

TYC\,4209 is a fairly bright EDD that has displayed significant changes over the last 
decade. Due to its fortuitous location in the Continuous Viewing Zone of the {\sl James Webb 
Space Telescope}, it will be possible to promptly monitor the aftermaths of possible 
future dust production events using mid-IR spectroscopy, which will provide a unique opportunity to study the 
associated mineralogical changes.

\section*{Acknowledgements}
We thank the referee, Inseok Song, for his insightful comments that improved the paper.
This work is based in part on observations made with the Spitzer 
Space Telescope, which was operated by the Jet Propulsion Laboratory, 
California Institute of Technology under a contract with NASA.
This publication makes use of data products from the Wide-field Infrared Survey 
Explorer, which is a joint project of the University of California, Los Angeles, 
and the Jet Propulsion Laboratory/California Institute of Technology, and NEOWISE, 
which is a project of the JetPropulsion Laboratory/California Institute of Technology. 
WISE and NEOWISE are funded by the National Aeronautics and Space Administration.
The publication makes use of data products from the Two Micron All Sky Survey, 
which is a joint project of the University of Massachusetts and the Infrared
Processing and Analysis Center/California Institute of Technology, funded by the 
National Aeronautics and Space Administration and the National Science Foundation.
This work has made use of data from the European Space Agency (ESA) mission
{\it Gaia} (\url{https://www.cosmos.esa.int/gaia}), processed by the {\it Gaia}
Data Processing and Analysis Consortium (DPAC,
\url{https://www.cosmos.esa.int/web/gaia/dpac/consortium}). Funding for the DPAC
has been provided by national institutions, in particular the institutions
participating in the {\it Gaia} Multilateral Agreement.
This research has made use of the NASA/ IPAC Infrared Science Archive, which is 
operated by the Jet Propulsion Laboratory, California Institute of Technology, 
under contract with the National Aeronautics and Space Administration.
We used the VizieR catalogue access tool and the Simbad object data base at CDS to 
gather data. 
This project has been supported by the GINOP-2.3.2-15-2016-00003, 2019-2.1.11-T\'ET-2019-00056, 
K-131508, and KKP-43986 \'Elvonal grants of the Hungarian National Research, Development and Innovation Office (NKFIH). 
Authors acknowledge the financial support of the Austrian-Hungarian Action Foundation 
(101\"ou13, 104\"ou2). 
G.C. is supported by the NAOJ ALMA Scientific Research grant code 2019-13B.
A.B., Zs.B. and \'A.S. are supported by the Lend\"ulet 
Program  of the Hungarian Academy of Sciences, project No. LP2018-7/2021 and 
the KKP-137523 'SeismoLab' \'Elvonal grant of the Hungarian Research, Development and 
Innovation Office (NKFIH).
Zs.B., L.K., and K.V. acknowledges the support by the J\'anos Bolyai Research Scholarship 
of the Hungarian Academy of Sciences. K.V. was supported by the Bolyai + \'UNKP programme.
L.K. acknowledges the financial support of the Hungarian National 
Research, Development and Innovation Office grant NKFIH PD-134784.

\section*{Data Availability}
The {\sl Spitzer} data used in this paper are publicly available at Spitzer Heritage Archive (\url{https://sha.ipac.caltech.edu/applications/Spitzer/SHA/}).




\input{ms.bbl}




\appendix




\bsp	
\label{lastpage}
\end{document}

%% file: tab1.tex
\begin{table}
	\centering
	\caption{Spitzer/IRAC Photometry of the TYC\,4209 system. 
The full table of all 111 observations is available in electronic format. 
From the left- to the right-hand side, the columns correspond to 
the Astronomical Observing Request (AOR) key of the observation, 
the Modified Julian Date of the 3.6\,{\micron} observation (computed as the mean
of the MJDs of the individual frames at the nine dither positions), 
the measured flux density and its uncertainty at 3.6\,{\micron}, 
the Modified Julian Date of the 4.5\,{\micron} observation, 
the measured flux density and its uncertainty at 4.5\,{\micron}.
\label{tab:iractable}}
	\begin{tabular}{lcccccc} 
	\hline
	AOR & MJD$_{\mathrm{3.6}}$ & $F_{\mathrm{3.6}}$ & $\sigma F_{\mathrm{3.6}}$ & 
	      MJD$_{\mathrm{4.5}}$ & $F_{\mathrm{4.5}}$ & $\sigma F_{\mathrm{4.5}}$ \\
	    &   (days)             &    (mJy)           &    (mJy)                  &
	        (days)             &    (mJy)           &    (mJy)                  \\       
	\hline
63183360 &  57935.93048 &   39.68 &    0.11 &  57935.92921 &   37.00 &    0.16 \\
63183616 &  57946.87000 &   39.56 &    0.11 &  57946.86873 &   36.67 &    0.13 \\
63183872 &  57956.23380 &   39.14 &    0.13 &  57956.23253 &   35.95 &    0.11 \\
63184128 &  57965.70466 &   39.41 &    0.10 &  57965.70339 &   36.34 &    0.09 \\
63184384 &  57975.09877 &   39.58 &    0.20 &  57975.09751 &   36.21 &    0.12 \\
63184640 &  57986.73201 &   39.01 &    0.11 &  57986.73074 &   35.66 &    0.10 \\
63184896 &  57995.04548 &   38.52 &    0.12 &  57995.04421 &   35.09 &    0.11 \\
63185152 &  58005.22230 &   38.89 &    0.19 &  58005.22103 &   35.42 &    0.07 \\
63185408 &  58016.35168 &   38.94 &    0.17 &  58016.35042 &   35.31 &    0.08 \\
63185664 &  58026.34130 &   38.79 &    0.14 &  58026.34004 &   35.53 &    0.19 \\
	\hline
	\end{tabular}
\end{table}